\documentclass{mn2e}

\usepackage{graphicx}

\newcommand{\msun}{\mathrm{M}_\odot}
\newcommand{\rsun}{\mathrm{R}_\odot}
\newcommand{\lsun}{\mathrm{L}_\odot}
\newcommand{\mtwodot}{\dot{M}_2}
\newcommand{\mtwodote}{\dot{M}_{2e}}
\newcommand{\K}{\hbox{K}}
\newcommand{\yr}{\hbox{yr}}

\newcommand{\araa}{ARA\&A}
\newcommand{\apss}{Ap\&SS}
\newcommand{\mnras}{MNRAS}
\newcommand{\aap}{A\&A}
\newcommand{\aaps}{A\&AS}
\newcommand{\apj}{ApJ}
\newcommand{\aj}{AJ}
\newcommand{\apjl}{ApJL}
\newcommand{\apjs}{ApJ Supp}
\newcommand{\pasp}{PASP}

\title[Mass Transfer between Double White Dwarfs]{Mass Transfer between Double White Dwarfs}

\author[T. R. Marsh, G. Nelemans,  D.Steeghs]
       {T. R. Marsh$^{1,2}$, G. Nelemans$^3$, D. Steeghs$^4$\\
        1. Department of Physics, University of Warwick, Coventry CV4 7AL, UK\\
        2. Department of Physics and Astronomy, University of Southampton,
Highfield, Southampton S017 1BJ, UK\\
        3. Institute of Astronomy, Madingley Road, Cambridge CB3 OHA, UK \\
        4. Harvard--Smithsonian Center for Astrophysics, 60 Garden Street, Cambridge, MA 02138, USA
}

\date{Accepted ;
      Received ;
      in original form}

\pagerange{\pageref{firstpage}--\pageref{lastpage}}
\pubyear{2001}

\begin{document}

\maketitle

\label{firstpage}

\begin{abstract}
Three periodically variable stars have recently been discovered
(V407~Vul, $P = 9.5\,\min$; ES~Cet, $P = 10.3\,\min$;
RX~J0806.3+1527, $P = 5.3\,\min$) with properties that suggest that
their photometric periods are also their orbital periods, making them
the most compact binary stars known. If true, this might indicate
that close, detached, double white dwarfs are able to survive the onset of mass
transfer caused by gravitational wave radiation and emerge as the
semi-detached, hydrogen-deficient stars known as the AM~CVn stars. The
accreting white dwarfs in such systems are large compared to the
orbital separations. This has two effects: first it makes it likely
that the mass transfer stream can hit the accretor directly, and
second it causes a loss of angular momentum from the orbit which can
destabilise the mass transfer unless the angular momentum lost to the
accretor can be transferred back to the orbit. The effect of the
destabilisation is to reduce the number of systems which survive mass
transfer by as much as one hundred-fold. In this paper we analyse this
destabilisation and the stabilising effect of a dissipative torque
between the accretor and the binary orbit.  We obtain analytic
criteria for the stability of both disc-fed and direct impact accretion,
and carry out numerical integrations to assess the importance of
secondary effects, the chief one being that otherwise stable systems
can exceed the Eddington accretion rate. We show that to have any
effect upon survival rates, the synchronising torque must act on a
timescale of order 1000 years or less. If synchronisation
torques are this strong, then they will play a significant role in the
spin rates of white dwarfs in cataclysmic variable stars as well.
\end{abstract}


\section{Introduction}
In recent years there have been discoveries of close detached double
white dwarfs at a rate that suggests that there may be of order 200
million such systems in our Galaxy, making them the largest population
of close binary stars (e.g. Marsh, Dhillon \& Duck 1995;  Napiwotzki et~al. 2003). These systems
were predicted theoretically long ago (Webbink 1979; Webbink 1984; Iben \& Tutukov 1984),
and were proposed as potential Type~Ia supernova progenitors, the
explosion being triggered by the action of gravitational wave losses
which cause them to merge. Only those systems with total mass in
excess of $1.4\,\msun$ have the potential to become Type~Ia
supernovae. These represent only a few percent of merging systems (Iben, Tutukov \& Yungelson 1997), 
which raises the question as to the fate of the less
massive systems. Possible outcomes which have been suggested are sdB
stars and R~CrB stars (Webbink 1984; Iben 1990;
Saio \& Jeffery 2002).  Another possibility is that the stars survive as binary
systems to become the semi-detached accreting binary stars known as
the AM~CVn stars (Nather, Robinson \& Stover 1981; Tutukov \& Yungelson 1996;
Nelemans et~al. 2001). These remarkable systems have extremely short
orbital periods (17 to 65 minutes) made possible because of the high
densities of their degenerate donor stars. Moreover they feature
accretion discs composed largely of helium which offer the opportunity
to see the effect of abundance upon disc physics.

In their study of the evolution of AM~CVn systems,
Nelemans et~al. (2001) considered the evolution of detached double
white dwarfs into the semi-detached phase, which requires the systems
to pass through an ultra-compact phase when orbital periods as low as
two or three minutes are possible. The stability of the mass transfer
during this stage is crucial as to whether the systems survive intact
or merge.  An additional complication is that even in stable systems the
equilibrium mass transfer rate can be (highly) super-Eddington,
resulting in mass and angular momentum loss and probably merging of
the system (Han \& Webbink 1999; Nelemans et~al. 2001). Mass transfer
between two white dwarfs has generally been taken to be stable if the
mass ratio $q = M_2/M_1$, where $M_1$ is the mass of the accretor, is
smaller than about $2/3$. However, Nelemans et~al.
identified the interesting possibility of the mass transfer stream
hitting the accreting white dwarf directly during this phase. They
realised that mass transfer without an accretion disc may destabilise
the mass transfer process for lower mass ratios still because the
normal transfer of angular momentum from disc to orbit would be
absent.

Assuming that unstable systems do not survive, Nelemans et~al. (2001)
found that the destabilisation by direct impact accretion could have a
dramatic effect upon the number of systems able to survive the onset
of mass transfer. If close white dwarfs typically do not survive this
phase, then it may mean that either evolution from helium stars
(Iben \& Tutukov 1991;  Nelemans et~al. 2001) or from immediate post
main-sequence donors (Podsiadlowski, Han \& Rappaport 2003) are more important
channels for the formation of AM~CVn stars than the double white dwarf
route.

There is now some observational evidence, that the very short orbital
periods suggested by the white dwarf merger model are possible. The
three stars V407~Vul (Cropper et~al. 1998), ES~Cet
(Warner \& Woudt 2002), and RX~J0806.3+1527 (Ramsay, Hakala \& Cropper 2002;
Israel et~al. 2002) show periodic variations on periods of $9.5$,
$10.3$ and $5.3\,\min$ respectively, and in each case no other period
is seen. In addition, there is evidence that each of these stars
appears hydrogen deficient (strongest for ES~Cet), as expected if
these periods are truly their orbital periods. Finally, V407~Vul and
RX~J0806.3+1527 show several observational characteristics compatible
with their still being in a phase of direct impact accretion
(Marsh \& Steeghs 2002; Ramsay et~al. 2002;
Israel et~al. 2002), or alternatively that they are still detached
but approaching mass transfer (Wu et~al. 2002;
Strohmayer 2002; Strohmayer 2003;
Hakala et~al. 2003). What is still lacking is spectroscopic proof
of the ultra-short periods of these stars, nevertheless it is likely
that to understand these systems we will need to understand mass
transfer between white dwarfs.

In this paper we study the stability of mass transfer in double white
dwarf binary stars. In particular we study whether, and under what
circumstances, dissipative coupling of the accretor to the orbit,
through either a magnetic field or tidal forces, can stabilise the
mass transfer. We find that the finite size of the accretor not only
leads to direct impact accretion, but also destabilises the accretion
whether it occurs by direct impact or through a disc. The
destabilisation has a significant effect upon the survival of double
white dwarfs as binary stars. 

We start in section~\ref{sec:setup} by setting up the equations
describing the onset of mass transfer. Then, in
sections~\ref{sec:analytic} and \ref{sec:numerical} we present analytical
and numerical solutions of these equations. In
section~\ref{sec:survive} we show the possible impact of our findings
for the population of AM~CVn systems while in
section~\ref{sec:discuss} we discuss the uncertainties and open
questions.

\section{The equations governing the evolution of mass transfer and white dwarf spin}
\label{sec:setup}

To consider the stability of mass transfer, we need a model that
includes a description for the variation of mass transfer rate with
the degree of over-filling of the Roche lobe by the donor star. We
discuss this in section~\ref{sec:mdot}. We also need to consider the question
of feedback of angular momentum from the accreting white dwarf to the
orbit (similar to Priedhorsky \& Verbunt (1988) who considered feedback
from a disc).  Before mass transfer starts, the two white dwarfs will
typically have been orbiting one another in ever decreasing circles
for many millions of years. Tidal or magnetic coupling will be acting
to synchronise their spins with their orbit. We assume that the donor
is always synchronised. Once mass transfer starts, the accretor will
start to be spun up by the addition of high angular momentum
material. What matters for stability is whether it is able to couple
strongly enough to the orbit for the added angular momentum to be fed
back into the orbit. This will be the new element that we add to the
formalism of Priedhorsky \& Verbunt (1988), along with a proper account of
the angular momentum of material in the inner disc in the disc-fed
case.

To understand how the mass transfer rate changes with time, we first
need to know how the orbital separation changes. This evolves because
of orbital angular momentum loss, and thus we start by considering the
evolution of angular momentum.

\subsection{Angular momentum loss}
The orbital angular momentum changes because of the action of gravitational
radiation, the loss of mass during mass transfer and the coupling between
the accretor's spin and the orbit. These can be written as:
\begin{equation}
 \dot{J}_\mathrm{orb} = \dot{J}_\mathrm{GR} + \sqrt{G M_1 R_h} \mtwodot + 
\frac{k M_1 R_1^2}{\tau_S} \omega . \label{eq:jdot}
\end{equation}
The first term on the right-hand side represents the change from gravitational
wave radiation. It is given by
\begin{equation}
\dot{J}_\mathrm{GR} = - \frac{32}{5}
\frac{G^3}{c^5} \frac{M_1 M_2 M}{a^4} 
J_\mathrm{orb},\label{eq:gr}
\end{equation}
where $M_1$ and $M_2$ are the masses of the two stars, $a$ is the
orbital separation and $M=M_1+M_2$ is the total mass of the two stars
(Landau \& Lifshitz 1975).
The second term on the right of Eq.~\ref{eq:jdot} represents the loss
of angular momentum from the orbit upon mass loss at a rate of $\mtwodot$. We adopt
Verbunt \& Rappaport's (1988) description 
of the angular momentum of the transferred matter in terms of the radius $R_h$ 
of the orbit around the accretor which has the same specific angular momentum as the
transferred mass. We use Eq.~13 of Verbunt \& Rappaport (1988) to calculate
$R_h$, accounting for their inverted definition of mass ratio compared
to ours. 

The third term of Eq.~\ref{eq:jdot} is a torque from dissipative
coupling, tidal or magnetic, which we parameterise in terms of the
synchronisation timescale $\tau_S$ of the accretor, with the torque
linearly proportional to the difference between the accretor's spin and the
orbital angular frequencies, $\omega = \Omega_s - \Omega_o$.  The term
$k M_1 R_1^2$ is the moment of inertia of the accretor, where $k \approx 0.2$ 
(Motz 1952); a more accurate approximation will be derived later.
We ignore the spin angular momentum of the synchronised donor.

The orbital angular momentum is given by
\begin{equation}
J_\mathrm{orb} = \left(\frac{Ga}{M}\right)^{1/2} M_1 
M_2,\label{eq:j}
\end{equation}
from which it follows that
\begin{equation}
\frac{\dot{J}_\mathrm{orb}}{J_\mathrm{orb}}
= \left(1 - q\right) \frac{\mtwodot}{M_2} +
\frac{1}{2} \frac{\dot{a}}{a}, \label{eq:jcon}
\end{equation}
for conservative mass transfer ($\dot{M} = 0$) which we assume to be the case throughout this paper.

Combining Eqs~\ref{eq:jdot}, \ref{eq:gr} and \ref{eq:j} one obtains
\begin{equation}
\frac{\dot{J}_\mathrm{orb}}{J_\mathrm{orb}} = 
\frac{\dot{J}_\mathrm{GR}}{J_\mathrm{orb}} +
\sqrt{(1+q)r_h} \frac{\mtwodot}{M_2}
+ \frac{k M_1R_1^2}{\tau_S J_\mathrm{orb}}\, \omega ,
\label{eq:jcomb}
\end{equation}
where $r_h = R_h/a$, and then Eqs~\ref{eq:jcon} and
\ref{eq:jcomb} can be combined to obtain an expression for the rate of change of the orbital separation:
\begin{equation}
\frac{\dot{a}}{2a} =
\frac{\dot{J}_\mathrm{GR}}{J_\mathrm{orb}}
+ \frac{k M_1R_1^2}{\tau_S J_\mathrm{orb}}\, \omega
- \left(1 - q - \sqrt{(1+q) r_h} \right) 
\frac{\mtwodot}{M_2} . \label{eq:adot}
\end{equation}
This equation shows that the orbital separation decreases due to
gravitational wave losses, increases due to the dissipative torque if the
accretor is spinning faster than the binary, and may decrease or
increase as a result of mass transfer, depending upon the mass
ratio. The last term is the reason why mass transfer can be unstable
because it can lead to a runaway where mass transfer causes the
separation to decrease which in turn causes the mass transfer rate to
increase.  To see this in detail, we now specify how the mass transfer rate
depends upon the degree to which the Roche lobe is over-filled, and how
the degree of overfill evolves with time.

\subsection{The mass transfer rate and the evolution of the over-fill factor}
\label{sec:mdot}

We define the over-fill of the Roche lobe to be
\begin{equation}
\Delta = R_2 - R_L,
\end{equation}
where $R_2$ is the radius of the donor and $R_L$ the radius of its
Roche lobe. The mass loss rate from the donor monotonically increases
with $\Delta$. How it does so has been investigated by many authors
(Paczy\'nski \& Sienkiewicz 1972; Webbink 1977; Savonije 1978;
Meyer \& Meyer-Hofmeister 1983; Ritter 1988;
Priedhorsky \& Verbunt 1988). These investigations divide into two
classes which we will term ``adiabatic'' and ``isothermal''. At low
mass transfer rates, the inner Lagrangian point lies close to the
photosphere of the donor, and we may adopt the formalism of
Ritter (1988) to give
\begin{equation}
\mtwodot = - \dot{M}_0 e^{\Delta/H}, \label{eq:mdot}
\end{equation}
where $\dot{M}_0$ depends upon the density in the photosphere and $H$,
the scale-height, is given by $H = k_B T_2 /\mu g_2$ where $k_B$ is
Boltzmann's constant, and $T_2$ and $g_2$ are the surface temperature
and gravity of the donor. We term this the ``isothermal'' model. It is appropriate
to cataclysmic variable stars and to AM~CVn stars at long orbital periods. At
high mass transfer rates, the inner Lagrangian point will move below
the photosphere, and the exponential relation of the isothermal case
will break down. The correct way to treat this is to compute the
stellar structure (Savonije 1978). However with all
calculations overshadowed by the uncertainty in the synchronisation
torque, we prefer to move to the opposite extreme of the isothermal
case, which is the adiabatic response, which for a white dwarf donor
gives a mass transfer rate of
\begin{equation}
\mtwodot = - f(M_1, M_2, a, R_2) \Delta^3, \label{eq:mdot_ad}
\end{equation}
for $\Delta > 0$, and zero for $\Delta < 0$
(Webbink 1984). Combining results from Webbink (1984),
Chandrasekhar (1967) and Webbink (1977), the function
$f$ is given by
\begin{eqnarray}
f(M_1, M_2, a, R_2) &=& \frac{8 \pi^3}{9} \left(\frac{5 G m_e}{h^2}
\right)^{3/2} \left(\mu'_e m_n\right)^{5/2}
\label{eq:f} \\
& & \times \frac{1}{P} 
\left(\frac{3 \mu M_2}{5 r_L R_2}\right)^{3/2} \left[a_2(a_2-1)\right]^{-1/2}, \nonumber
\end{eqnarray}
where $m_e$ is the mass of an electron, $m_n$ is the mass of a nucleon,
$\mu'_e$ is the mean number of nucleons per free electron in the outer layers of
the donor (which we will assume to be $2$), $P$ is the orbital period,
$r_L = R_L/a$ and $\mu$ and $a_2$ are parameters associated with the
Roche potential (Webbink 1977):
\begin{eqnarray}
\mu &=& \frac{M_2}{M_1 + M_2},\\
a_2 &=& \frac{\mu}{x_{L1}^3} + \frac{(1-\mu)}{(1-x_{L1})^3},
\end{eqnarray}
where $x_{L1}$ is the distance of the inner Lagrangian point from the centre of the donor
in units of $a$  and we have swapped the order of the two masses 
relative to that of Webbink (1977) since in our case the secondary star is the donor.

The key difference between Eqs.~\ref{eq:mdot} and \ref{eq:mdot_ad} is
their sensitivity to changes in $\Delta$ which has an effect upon when
and how mass transfer is unstable, but not the rate at which mass
transfer proceeds when it is stable. Which approximation applies depends
upon the mass transfer rate, with the dividing line given by the mass transfer
rate sustainable under the isothermal model when the inner Lagrangian point is 
located at the photosphere. If the mass transfer rates during merging are much 
higher than this, then the adiabatic rate is the more suitable. Following 
Ritter (1988), the isothermal mass transfer rate scales as 
\begin{equation}
-\mtwodot \propto R_2^3 M_2^{-1} \left(\frac{T_2}{\mu}\right)^{3/2} \rho_\mathrm{ph},
\end{equation}
where we have left out a weak function of mass ratio. Here $R_2$,
$M_2$ and $T_2$ are the radius, mass and photospheric temperature of
the donor star while $\rho_\mathrm{ph}$ is the photospheric density
and $\mu$ is the mean molecular mass.  For a main-sequence star of
$M_2 = 0.5\,\msun$, Ritter (1988) quotes $R_2 = 0.52\,\rsun$,
$T_2 = 3520\,\K$, $\rho_\mathrm{ph} = 1.6 \times 10^{-5}\,{\rm
g}\,{\rm cm}^{-3}$, $\mu = 1.33$ and $-\mtwodot = 0.9 \times
10^{-8}\,\msun\,\yr^{-1}$.  We scale from this to a white dwarf of $M_2 =
0.5\msun$ for which $R_2 = 0.013\,\rsun$. Taking
Koester's (1980) $T_\mathrm{eff} =
20$,$000\,\K$, $\log g = 8$ model of a DB (helium dominated)
atmosphere, we find that at the photosphere $\mu = 3.5$ and
$\rho_\mathrm{ph} = 2.8 \times 10^{-5} \,{\rm g}\,{\rm cm}^{-3}$. From
these figures we obtain $-\mtwodot = 0.8 \times 10^{-12}
\,\msun\,\yr^{-1}$. This rate is very much lower than typical
equilibrium accretion rates at contact which are typically of order
$10^{-5} \, \msun\,\yr$, except in the case of implausibly low
component masses. Therefore the adiabatic approximation is the more
appropriate.  Hence for the remainder of this paper, unless stated 
otherwise, we consider the adiabatic model of mass transfer.

The mass transfer rate contributes towards the evolution of the
over-fill factor which changes at a rate given by
\begin{eqnarray}
\frac{d \Delta}{d t} &=& \left( R_2 \zeta_2 - R_L \zeta_{r_L}\right)
                      \frac{\mtwodot}{M_2} - R_L \frac{\dot{a}}{a},\\
                      &=& R_2 \left[ \left(\zeta_2 - \zeta_{r_L}\right)
                      \frac{\mtwodot}{M_2} - \frac{\dot{a}}{a}\right]
                      + \Delta \left( \zeta_{r_L} \frac{\mtwodot}{M_2} +
                      \frac{\dot{a}}{a} \right), \label{eq:ddel}
\end{eqnarray}
where 
\begin{eqnarray}
\zeta_2 &=& \frac{d \log R_2}{d \log M_2},\\
\zeta_{r_L} &=& \frac{d \log (R_L/a)}{d \log M_2}.
\end{eqnarray}
The factor $\zeta_{r_L} = 1/3$ for Paczy\'nski's
(1971) small $q$ approximation for the Roche
lobe radius while for Eggleton's
(1983) approximation it is given by
\begin{equation}
\zeta_{r_L} = \frac{(1+q)}{3} \frac{2\ln (1+q^{1/3}) - q^{1/3}/(1+q^{1/3})}
{0.6q^{2/3} + \ln ( 1 + q^{1/3} )} \approx \frac{1}{3},
\end{equation}
for conservative mass transfer, $\dot{M} = 0$.
White dwarfs typically become larger as their mass decreases, and
so $\zeta_2$ is negative, with typical values of $-0.6 < \zeta_2 < -0.3$.

Neglecting the second term of Eq.~\ref{eq:ddel} since $|\Delta| \ll R_2$,
and substituting for $\dot{a}/a$ from Eq.~\ref{eq:adot} gives
\begin{eqnarray}
\frac{1}{2R_2} \frac{d \Delta}{d t} &=& -\frac{\dot{J}_\mathrm{GR}}{J_\mathrm{orb}}
- \frac{k M_1 R_1^2}{\tau_S J_\mathrm{orb}}\, \omega \, + \label{eq:deldot}
\\ \nonumber 
& & \left(1 + \frac{\zeta_2-\zeta_{r_L}}{2} - q - \sqrt{(1+q) r_h} \right) \frac{\mtwodot}{M_2}.
\end{eqnarray} 
This equation, which is essentially identical to Eq.~20 of
Priedhorsky \& Verbunt (1988), shows that the over-fill grows due to
gravitational wave losses and shrinks from dissipative coupling, if
the accretor is spinning faster than the binary. It is the last term
that can lead to instability, for if the bracketed part is negative
then one has a situation where mass transfer ($\mtwodot < 0$) causes
$\Delta$ to grow, which increases the mass transfer rate, until the
whole process runs away. The destabilisation caused by direct impact
is contained in the term $\sqrt{(1+q) r_h}$ representing the loss of
angular momentum from the orbit. In circumstances that we will elucidate,
this instability can be counteracted by transferring angular momentum 
back to the orbit via the term in $\omega$.

In the case of disc accretion, it is usual to assume that all the
angular momentum contained in the mass gained from the donor is fed back to the orbit, which
means that the terms involving $\omega$ and $\sqrt{(1+q) r_h}$ do not
appear.  In this case, mass transfer is stable provided that
\begin{equation}
q < 1 + \frac{\zeta_2-\zeta_{r_L}}{2},
\end{equation}
because as the mass transfer rate increases and $\mtwodot$ gets more
negative, the last term in Eq.~\ref{eq:deldot} balances the first so
that $\dot{\Delta} = 0$ and $\ddot{M}_2 = 0$. For a white dwarf donor
with $\zeta_2 \approx -1/3$ and for $\zeta_{r_L} \approx 1/3$, this translates to the
widely-used ``$q < 2/3$'' for stability.  Relative to this limit, the
$\sqrt{(1+q) r_h}$ term causes instability. However, even with a disc,
the material at the inner edge of the disc still has some angular
momentum. In the very tight binary stars we consider here, this turns
out to be very significant, as we will see in
section~\ref{sec:disc}.  Therefore, whether or not accretion is
through a disc, we need to include the $\omega$ term, and
therefore to understand how the accretor's spin rate evolves with time.

\subsection{The Evolution of the Accretor's Spin Rate}

The accretor will spin up as a result of the addition of mass, and
spin down because of the dissipative torque, if the accretor spins 
faster than the binary. Conserving angular momentum one obtains
\begin{eqnarray}
\frac{d \Omega_s}{d t} &=& 
\frac{d\Omega_o}{ d t} + \frac{d \omega}{d t} \nonumber \\
&=&  \left( \lambda \Omega_s  - 
\frac{\sqrt{G M_1 R_h}}{k R_1^2}\right) \frac{\mtwodot}{M_1}
- \frac{\omega}{\tau_s},
\label{eq:omdot}
\end{eqnarray}
where 
\begin{equation}
\lambda = 1 + 2 \frac{d \log R_1}{d \log M_1} + \frac{d \log k}{d \log M_1} ,
\end{equation}
a factor which arises from the change in $k M_1 R_1^2$
with the addition of mass.

We mentioned earlier that the moment of inertia factor $k$ is
approximately $0.2$, but more accurately it is a function of mass, and
decreases at high masses. To allow for this we computed values of $k$
as a function of mass from Chandrasekhar's equation of state for
zero-temperature degenerate matter (Chandrasekhar 1967). We
fitted these with the function
\begin{equation}
k = 0.1939 \left(1.44885 - M_1\right)^{0.1917},
\end{equation}
which fits to better than $2$ per cent over the allowable mass range. It should be noted however, that although
this function is superior to a constant, we are not entirely self-consistent since we employ Eggleton's
zero temperature mass--radius relation, quoted by Verbunt \& Rappaport (1988):
\begin{eqnarray}
\frac{R}{\rsun} &=& 0.0114 \left[ \left(\frac{M}{M_\mathrm{Ch}}\right)^{-2/3} -
\left(\frac{M}{M_\mathrm{Ch}}\right)^{2/3} \right]^{1/2} \label{eq:egg_mr}\\
& & \times \left[ 1 + 3.5 \left(\frac{M}{M_p}\right)^{-2/3} + \left(\frac{M}{M_p}\right)^{-1}\right]^{-2/3} ,\nonumber
\end{eqnarray}
where $M_\mathrm{Ch} = 1.44 \,\msun$ and $M_p = 0.00057\,\msun$.  The advantage
of this relation is that it provides us with a single relation that matches
Nauenberg's (1972) high-mass relation but
also allows for the change to a constant density configuration at low masses
(Zapolsky \& Salpeter 1969). Eggleton's relation applies for the whole range $0 < M <
M_\mathrm{Ch}$ which means that we can use a single relation for both white
dwarfs, without discontinuities. Our relation for $k$ does not account for the
effects included by Zapolsky \& Salpeter (1969), and therefore it is probably an
underestimate of $k$ at low masses (since $k = 0.4$ for constant density
spheres). The denominator in Eq.~\ref{eq:egg_mr} makes at least a 10 per cent
difference to the radius for $M < 0.07 \,\msun$. However, the lowest mass stars
that can evolve off the main-sequence within a Hubble time have helium core
masses of $\sim 0.1\,\msun$, and so it is hard to see how white dwarfs of lower
mass can be formed.  Therefore we don't think that the extent of our
approximation for $k$ will be a problem.

Eqs~\ref{eq:deldot} and \ref{eq:omdot} describe the evolution of the
overfill and the spin rate of the accretor, but need modification when
accretion switches from direct impact to disc mode and when the donor reaches
its break-up rate.  We have already mentioned that it is usual to
assume that all the angular momentum is fed back into the orbit when a
disc forms (Priedhorsky \& Verbunt 1988), but because in our case the
accretor is relatively large compared to the binary, in the next
section we show how to account for the significant angular momentum
lost at the inner edge of the disc.

\subsection{Disc formation}
\label{sec:disc}
As mass transfer proceeds between two white dwarfs, the mass ratio become
smaller and the orbital separation increases. Both of these changes mean that
there will come a point when the stream no longer hits the accretor directly and
an accretion disc will form. This occurs when the minimum radius reached by the
stream, $R_\mathrm{min}$, exceeds the radius of the primary star, $R_1$.  To
calculate when this occurs we use Eq.~6 of Nelemans et~al. (2001). On formation
of a disc, most of the angular momentum of the accreted material will be
transferred back to the orbit by tidal forces at the outer edge of the disc
(Priedhorsky \& Verbunt 1988).  However, material at the inner edge of the disc still
has angular momentum (e.g. disc accretion acts to spin up the accretor). This
can be accommodated within Eqs~\ref{eq:deldot} and \ref{eq:omdot} simply by
replacing $R_h$ by $R_1$ and $r_h= R_h/a$ by $r_1 = R_1/a$ whenever $R_1 <
R_\mathrm{min}$. This is because $R_h$ was defined as the radius of a keplerian
orbit having the same angular momentum as the stream, whereas the inner disc has
the angular momentum of a keplerian orbit of radius $R_1$.

Since it is always the case that $R_h > R_\mathrm{min}$, when a
disc starts to form we must have $R_h > R_1$, so the switch corresponds to a
decrease in the angular momentum lost from the orbit as a consequence of mass transfer,
which therefore causes the mass transfer rate to drop. It is important to 
realise however that $R_1$ is not that much smaller than $R_h$ and so the
destabilisation that affects direct impact accretion affects disc accretion
to a very large extent as well. Moreover, the square root dependence upon $r_1$ 
suggests that this effect may not be negligible in wider binary systems, a point
we return to in section~\ref{sec:cvs}.

The change from $R_h$ to $R_1$ is the only one we make once a disc has
formed; we assume that the spin--orbit coupling expressed through the
synchronisation timescale remains unchanged.

\subsection{Maximum spin rate of the accretor}
\label{sec:break}
For low synchronisation torques, the accretor may be spun up to
its break-up rate at which point spin--orbit coupling will presumably
strengthen markedly through bar-mode type instabilities or the
shedding of mass. We model this as follows:
during evolution if $\Omega_s = \Omega_k$, where
\begin{equation}
\Omega_k = \sqrt{\frac{GM_1}{R_1^3}}
\end{equation}
is the keplerian orbital angular frequency at the surface of the accretor,
and $\dot{\Omega}_s > \dot{\Omega}_k$, where
\begin{equation}
\dot{\Omega}_k = - \frac{\Omega_k}{2} \left(1 - 3\zeta_1\right)\frac{\mtwodot}{M_1},
\end{equation}
and
\begin{equation}
\zeta_1 = \frac{d \log R_1}{d \log M_1},
\end{equation}
then we force $\dot{\Omega}_s = \dot{\Omega}_k$ by alteration of the synchronisation
timescale to a new value $\tau'_s$ given by
\begin{equation}
\frac{\omega}{\tau'_s} = \left( \lambda \Omega_k  - 
\frac{\sqrt{G M_1 R_h}}{k R_1^2}\right) \frac{\mtwodot}{M_1} - \dot{\Omega}_k. \label{eq:newtau}
\end{equation}
This revised value is then used in place of $\omega/\tau_s$ in
Eq.~\ref{eq:deldot} for the evolution of the over-fill factor to give
\begin{eqnarray}
\frac{1}{2R_2} \frac{d \Delta}{d t} &=& 
-\frac{\dot{J}_\mathrm{GR}}{J_\mathrm{orb}}
+ \left(1 + \frac{\zeta_2-\zeta_{r_L}}{2} - q - \right. \label{eq:deldotalt} \\ \nonumber
& & \left. k \lambda \sqrt{(1+q) r_1} \right) \frac{\mtwodot}{M_2} + 
\frac{k M_1 R_1^2 \dot{\Omega}_k}{J_\mathrm{orb}},
\end{eqnarray} 
where $r_1 = R_1/a$. This modification clamps the spin rate of the accretor to
the keplerian spin rate at its surface and increases the rate at which angular
momentum is injected back into the orbit.

Eqs~\ref{eq:deldot} and \ref{eq:omdot} together with the above
modifications for disc formation and the maximum spin of the accretor can be
integrated to find how the mass transfer rate and the spin frequency
of the accretor evolve with time. Before looking at examples of this,
we first look at quasi-static solutions of the equations, with our main
objective being to uncover the conditions under which dissipative coupling
can stabilise direct impact accretion.

\section{Analytic Results}
\label{sec:analytic}
The strong dependence of $\mtwodot$ upon the overfill factor means that the
timescale of variations in mass transfer is much shorter
than the mass-transfer ($\equiv$GR) timescale. Therefore, quasi-steady-state
solutions, in which the evolution of binary parameters is neglected,
provide a very useful way to understand what happens during mass
transfer. To do so we assume that the coefficients multiplying
$\omega$ and $\mtwodot$ in Eqs~\ref{eq:deldot} and \ref{eq:omdot} do not change with time and
also that $\dot{\Omega}_o = \dot{\Omega}_k = 0$.  In
section~\ref{sec:numerical} we will describe the results of numerical
integrations where we do not make these assumptions.

The details of the analysis of quasi-static solutions are contained in the appendix.
We now describe the qualitative outcomes of this work, starting with the 
circumstances under which dynamical instability is or is not guaranteed to
occur, and then moving onto the main focus of this paper, the stabilising effect
of dissipative spin--orbit coupling.

\begin{figure}
\hspace*{\fill}
\includegraphics[angle=270,width=\columnwidth]{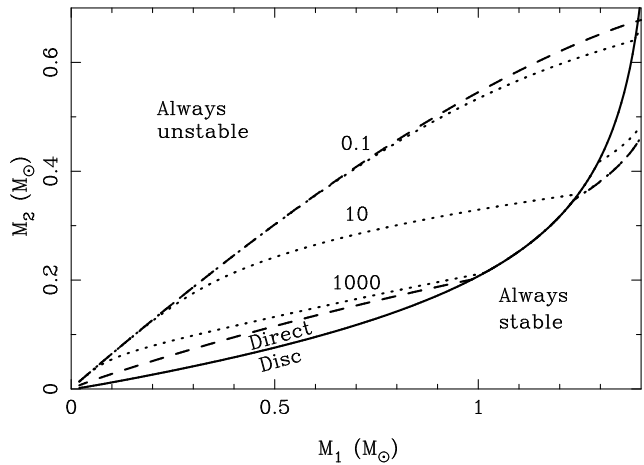}
\hspace*{\fill}
\caption{The upper dashed line shows the dynamical stability limit 
(Eq.~\protect\ref{eq:unstable}), while the lower dashed
line shows Nelemans et al.'s stricter criterion
(Eq.~\protect\ref{eq:stable}), accounting for the switch between direct impact
and disc accretion at $M_1 \approx 1 \, \msun$. The solid line shows the transition between disc and direct impact 
accretion. The three dotted lines show how Nelemans et al.'s strict stability limit
is relaxed when dissipative torques feed angular momentum from the accretor back
to the orbit (Eq.~\protect\ref{eq:sync}), once again accounting for both the direct
impact and disc accretion cases. The three lines are labelled by the synchronisation 
timescale in years. the line for $\tau_S = 1000\,\yr$ coincides with the lower dashed
line for $M_1 > 1.2\,\msun$ giving the dash-dot line.}
\label{fig:param}
\end{figure}

\subsection{Dynamically stable and unstable solutions}
\label{sec:equilibrium}

Depending upon the binary parameters, there are three
possible outcomes: (i) guaranteed dynamical instability, (ii)
guaranteed stability, and (iii) the intermediate case of either
stability or instability, depending upon the degree of spin--orbit
coupling. It is the third case that is of most interest in this paper;
we leave this to the next sub-section. Here we summarise the first two
cases.

Mass transfer is dynamically unstable and will lead to a merger, regardless
of synchronisation torques, if 
\begin{equation}
q > 1 + \frac{\zeta_2-\zeta_{r_L}}{2} - k r_1^2 (1+q) \lambda, \label{eq:unstable}
\end{equation}
(Eq.~\ref{eq:unstable_app}). This is nothing more
than the usual ``$q > 2/3$'' condition for dynamical instability,
very slightly modified by the last term which accounts for the moment of
inertia of the accretor. No amount of spin--orbit coupling can stabilise
the mass transfer in this case. This is the case studied by 
Saio \& Jeffery (2002) amongst others; it is also probably the most common outcome
of a merger as observed double white dwarfs have fairly equal masses
(Maxted, Marsh \& Moran 2002).

In the case of direct impact accretion, stable mass transfer is guaranteed, 
regardless of spin--orbit coupling, if
\begin{equation}
q < 1 + \frac{\zeta_2-\zeta_{r_L}}{2} - \sqrt{(1+q) r_h},\label{eq:stable}
\end{equation}
(Eq.~\ref{eq:stable_app}). The same equation with $r_h$ replaced by $r_1$ gives
the limit in the case of disc-fed accretion. Eq.~\ref{eq:stable} is exactly that
derived by Nelemans et~al. (2001) for the case of no feedback of the angular
momentum of the accreted material to the orbit. The third term on the right-hand
side of this equation is significantly larger in magnitude than its counterpart
in Eq.~\ref{eq:unstable} as can be seen in Fig.~\ref{fig:param} which shows
parameter constraints for two stars obeying Eggleton's (zero temperature)
mass-radius relation, Eq.~\ref{eq:egg_mr}.  In this figure, the upper dashed
line shows the limit, Eq.~\ref{eq:unstable}, for guaranteed instability, while
the lower dashed line shows the limit, Eq.~\ref{eq:stable}, below which mass
transfer is stable, accounting for whether the accretion is def by the disc or by
direct impact. In between these two lines, is a zone in which mass transfer is
stable or unstable according to the strength of spin--orbit coupling. This zone
is very significant, both in the fraction of parameter space it occupies, and
even more so when one accounts for typical parameters at birth
(section~\ref{sec:survive}).  We therefore devote a separate section to it, with
details once again left to the appendix.

\subsection{Stability of Equilibrium}

The boundaries of guaranteed stability and instability are defined by the
existence (or not) of quasi-static solutions of Eqs~\ref{eq:deldot} and 
\ref{eq:omdot}. The intermediate region of parameter space adds the new
possibility of unstable quasi-static solutions.
A linear stability analysis leads to the following condition (Eq.~\ref{eq:sync_app})
for these solutions to be stable:
\begin{equation}  
\frac{1}{\tau_S} > \left[ \beta \left(1 + \frac{\zeta_2-\zeta_{r_L}}{2} - q - \sqrt{(1+q) r_h} \right)
 +  \lambda q \right] \frac{\mtwodote}{M_2},
\label{eq:sync} 
\end{equation}
where $-\mtwodote$ is the equilibrium mass transfer rate and $\beta = 6 R_2 /
\Delta_e$ where $\Delta_e$ is the equilibrium value of the overfill factor
corresponding to $\mtwodote$ (see appendix for details). $\beta$ is a
dimensionless factor, typically of order $10^3$ to $10^4$, which measures the
sensitivity of the mass transfer rate to changes in the overfill factor.  This
is where the dependence of the mass transfer rate upon $\Delta$ matters: the
more sensitive it is, the larger $\beta$ is, and, since both the term in
brackets multiplying $\beta$ and $\mtwodote$ are negative, the stronger the
synchronisation torque has to be to ensure stability.

The condition of Eq.~\ref{eq:sync} is a generalisation of Nelemans et
al.'s (2001) strict condition for stability (Eq.~\ref{eq:stable}) and
is the key result of this paper. Once more, this condition applies for
direct impact accretion, while for disc-fed accretion the $r_h$ must
be replaced by $r_1 = R_1/a$. This condition quantifies one's
expectation that spin--orbit coupling will stabilise mass transfer,
and essentially says that the synchronisation timescale must be less
than the timescale upon which the mass transfer rate can vary
significantly.  This is what one expects: the spin of the white dwarf
must be able to respond to variations of the mass transfer rate to
ensure stability.

\begin{figure}
\hspace*{\fill}
\includegraphics[angle=270,width=\columnwidth]{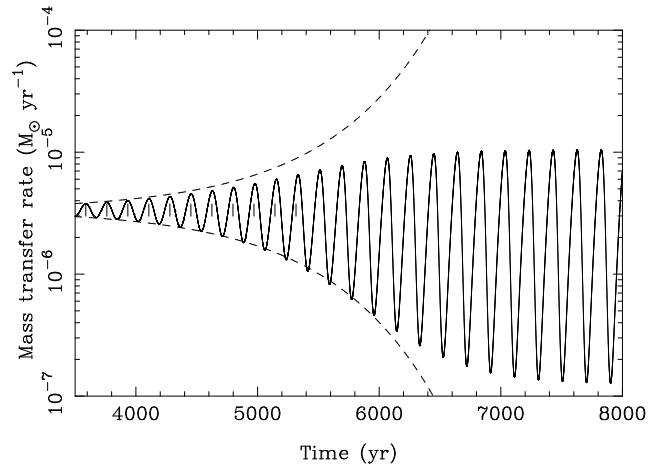}
\hspace*{\fill}
\caption{The evolution of mass transfer rate is shown (solid line)
for a marginally unstable case of $M_1 = 0.5\,\msun$ and $M_2 = 0.21 \,\msun$ and a
synchronisation timescale of $30\,\yr$. The model was started close to equilibrium 
and evolved with the masses and orbital separation held fixed. The short vertical lines mark the oscillation period
predicted from a linear stability analysis while the dashed curves represent the predicted amplitude.}
\label{fig:unstable}
\end{figure}

An example of marginally unstable mass transfer is shown in 
Fig.~\ref{fig:unstable} in which we also compare a
numerical integration (starting from a slight perturbation of the
equilibrium mass transfer rate) with the predictions of the linear
stability analysis. The oscillations result as first the white dwarf
is spun up, leading to injection of angular momentum back into the
orbit which reduces the transfer rate causing the white dwarf to spin
down, and so on. When the synchronisation torque is too weak, the
white dwarf does not respond fast enough to alterations of the
accretion rate to damp out perturbations, and their amplitude grows. 
In this particular case the amplitude saturates, but this is not of
great significance since it is only for rather finely-tuned cases 
that one does not have either stability or such a violent instability
that merger is inevitable. Moreover, long-term oscillations will not 
occur in practice because of the evolution of the component masses and orbital
separation which is not included in Fig.~\ref{fig:unstable}. 

\begin{figure}
\hspace*{\fill}
\includegraphics[angle=270,width=\columnwidth]{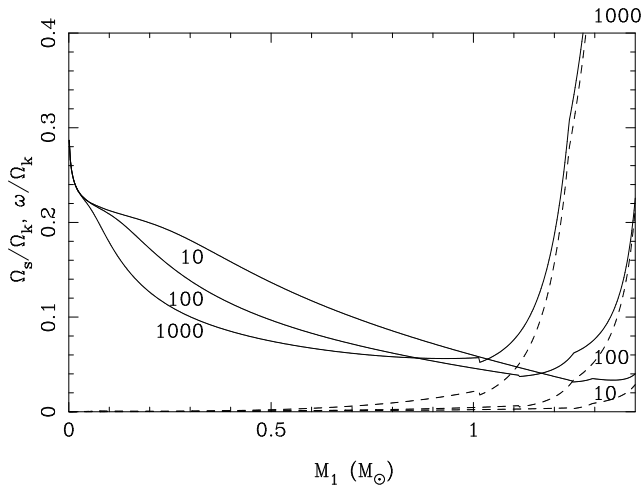}
\hspace*{\fill}
\caption{The equilibrium spin rate of the accretor $\Omega_s$ relative
to the keplerian angular velocity at its surface $\Omega_k$ is plotted
(solid curves) for marginal mass transfer stability for three different synchronisation
timescales (in years). The dashed curves show the differential spin rate
$\omega = \Omega_s - \Omega_o$ relative to $\Omega_k$.}
\label{fig:spin}
\end{figure}

To apply the stability criterion, Eq.~\ref{eq:sync}, one must first
calculate $\mtwodote$, which also gives
$\Delta_e$ (through Eq.~\ref{eq:mdot_ad}); this calculation is detailed
in the appendix.  Lines of stability for various synchronisation
timescales are plotted in Fig.~\ref{fig:param}. These show how the
action of dissipative torques expands the region of stable mass
transfer in the case of direct impact. The lower dashed line, marking
the onset of instability in the absence of any synchronising torque,
is raised to become one of the dotted lines (labelled by the
synchronisation timescale at the start of mass transfer). Clearly, the
synchronisation timescale must be short to have much effect, except
for very low mass systems. Nelemans et~al.'s
(2001) criterion (Eq.~\ref{eq:stable}) is the
limiting case for $\tau_S \rightarrow \infty$, while the standard
dynamical stability limit (Eq.~\ref{eq:unstable}) is the limit as
$\tau_S \rightarrow 0$.  

For large accretor masses, direct impact can be avoided for a wide range of
donor masses as the accretor becomes very small. The switch from $r_h$ to $r_1$
stabilises the mass transfer, and for a while stability becomes a case of
whether the accretion occurs through direct impact or not, with the dotted
stability lines following the disc accretion limit (solid line) in
Fig.~\ref{fig:param}.  However, at very high accretor masses ($M_1 >
1.2\,\msun$), even the switch to $r_1$ is insufficient, and the lines of
stability drop below the solid line marking the disc/direct transition. Thus
there are even regimes of disc accretion which are destabilised by the loss of
angular momentum from the inner disc.

Fig.~\ref{fig:spin} shows the equilibrium angular velocity of the accretor 
relative to the Keplerian angular velocity at its surface
in the case of systems just at the stability limit, the fastest
case. This figure shows that for synchronisation timescales of
interest for evolution, the accretor does not approach break-up. This
figure may appear counter-intuitive in that weaker synchronisation
causes slower rotation in some cases. This results from the higher
donor masses made possible by stronger synchronisation which lead to
much smaller orbits and orbital periods: for much of Fig.~\ref{fig:spin} the
accretor is almost synchronous with the orbit as can can be seen from the small
values of the differential spin rate, $\omega = \Omega_s - \Omega_o$ (dashed lines),
and really we are just seeing that the accretor can be close to filling its Roche lobe when
it is of low mass, and therefore by definition it rotates at a rate of the same
order of magnitude as the break-up rate.
Discontinuities at high masses are caused by the transition to disc accretion.

\subsection{Super-Eddington accretion}

Eqs.~\ref{eq:mdnorm} and \ref{eq:mdhigh} of the appendix give the 
equilibrium mass transfer rates for dynamically stable systems. Each of
these is of the form 
\begin{equation}
 \frac{-\mtwodote}{M_2} =  f(q) \frac{-\dot{J}_\mathrm{GR}}{J_\mathrm{orb}},
\end{equation}
where $f(q) \rightarrow \infty$ as $q$ increases towards the instability
limit. Given this, and that for massive systems the GR timescale can
be short, there are almost inevitably ranges of parameter space which, although stable,
lead to super-Eddington accretion. Han \& Webbink (1999) discuss this case
extensively, focussing upon the fraction of mass that can be
ejected. Although ejection of mass is needed for a system to survive
super-Eddington accretion, Han \& Webbink (1999) argue that it is likely
that the ejected mass will form a common-envelope around the binary
system which will lead to the merger of the two white dwarfs. Thus we
assume, as do Nelemans et~al. (2001), that the ultimate consequence of
super-Eddington accretion is merging and the loss of the system as a
potential AM~CVn progenitor. We are in effect assuming that dynamical
instability and super-Eddington accretion lead to the same end result:
merging.

In order to calculate the Eddington accretion rate, we follow
Han \& Webbink (1999) and use the difference in the Roche potential at the
inner Lagrangian point and the accretor, $\phi_{L1} - \phi_a$, to give the
energy released per unit mass. The Eddington accretion rate $\dot{M}_{\rm Edd}$ is then
given by
\begin{equation}
\dot{M}_{\rm Edd} = \frac{8\pi G m_p c M_1}{\sigma_T \left(\phi_{L1} - \phi_a\right)},
\end{equation}
where $\sigma_T$ is the Thomson cross-section of the electron and
$m_p$ the mass of a proton. This expression comes from taking two
proton masses per free electron, as appropriate for fully-ionised
helium or carbon. This value is increased over the usual solar-composition
limit because of the higher mass per free electron, but also because of the
relatively deep potential at the inner Lagrangian point in a double white
dwarf system. We consider a refinement of this expression in 
section~\ref{sec:method}. 

\section{Numerical integrations}
\label{sec:numerical}
The analytical results above take no account of how the system reaches
equilibrium or of the evolution of system parameters that occurs
during this process. For instance, they take no account of the
expected lengthening of the synchronisation timescale as the binary
separation increases, which may de-stabilise the mass
transfer. Similarly, the analytic results do not include the
possibility of the accretor reaching its break-up spin rate, which
increases coupling, and may stabilise mass transfer.  Finally, even
though equilibrium mass transfer rates can be calculated, it is the
maximum mass transfer rate that is of more interest from the point of
view of surviving contact. The significance of these possibilities
can only be answered through numerical integration.

\subsection{Method}

\label{sec:method}

We carried out fifth-order Runge-Kutta integrations, adapting the time-step as
the integrations proceeded. The integrations were started just before contact.
To simulate the effect of a long interval prior to contact, we started the
accretor with a differential spin rate of
\begin{equation}
\omega = - \tau_s \frac{d \Omega_o}{d t} .
\end{equation}
This ensures, through Eq.~\ref{eq:omdot}, that $d \omega / d t \approx
0$ in the case of strong coupling, as one expects, with the primary
star lagging slightly behind the increasing orbital frequency. For
large $\tau_s$, this can give $\Omega_s < 0$, in which case we fixed
its initial value at zero.

One of the purposes of the numerical integrations was to see if
systems could be stable and yet exceed the Eddington luminosity.  As
described above, Han \& Webbink (1999) have already detailed the
computation of the Eddington rate in double white dwarf binary
stars. However, they implicitly assume that the accretor co-rotates
with the binary orbit. Since we are explicitly allowing this not to be
the case, we need to correct Han \& Webbink's expression
$\phi_{L1} - \phi_a$. Instead we use the following formula
\begin{equation}
\phi_{L1} - \phi_a - \frac{1}{2} \mathbf{v}_i^2 + 
\frac{1}{2} \left(\mathbf{v}_i - \mathbf{v}_\omega\right)^2,
\end{equation}
where $\mathbf{v}_i$ is the impact velocity and $\mathbf{v}_\omega$ is
the velocity of the accretor at the point of impact, both measured in
the rotating frame. This expression comes from first subtracting the
contribution of the impact assuming that the accretor is co-rotating,
and then adding back in the true amount accounting for $\omega \neq
0$.  If $\omega = 0$, then $\mathbf{v}_\omega = \mathbf{0}$ and
Han \& Webbink's expression is returned as expected.  The
modified formula works equally for direct impact and disc accretion.
We implemented it by pre-computing a set of impact velocities and
locations which we interpolated during the integrations. In a
similar manner we also calculated the luminosity of any stream/disc
impact in the case of disc accretion. We found that once the accretor
was spinning at its break-up limit, the accretion luminosity was roughly 
cut in half relative to Han \& Webbink's (1999)
value, which essentially is the result of the virial theorem. However,
we also found that immediately after mass transfer starts, with the accretor 
rotating slowly, Han \& Webbink's (1999) value is 
a good approximation, and it is only this early phase that matters as far
as super-Eddington accretion is concerned.

\subsection{Scaling of the synchronisation timescale}
In the analytical calculations, we assumed that the synchronisation
timescale is fixed, whereas one expects it to increase as the separation of
the binary increases. For instance, Campbell (1983) calculates
the synchronisation due to dissipation of induced electrical currents
within the donor when the accretor is magnetic. The synchronisation
timescale in this case varies with the degree of asynchronism
$\omega$, but in the small $\omega$ limit Campbell (1983) finds
\begin{equation}
\tau_S \sim 2 \times 10^6 M_1 R_1^{-4} R_2^{-5} a^6 B^{-2} \,\yr ,
\end{equation}
where $B$ is the surface field of the accretor in Gauss and the other
quantities are in solar units. Putting $M_1 = 0.6\,\msun$, $R_1 =
0.012\,\rsun$, $R_2 = 0.02\,\rsun$, $a = 0.06\,\rsun$ and $B =
10^7\,\mathrm{G}$ gives $\tau_S \approx 10\,\yr$. However, this
ignores the high conductivity of degenerate gas which reduces the
dissipation and lengthens the timescale considerably for a fixed field
strength.  
\begin{figure*}
\hspace*{\fill}
\includegraphics[angle=270,width=0.8\textwidth]{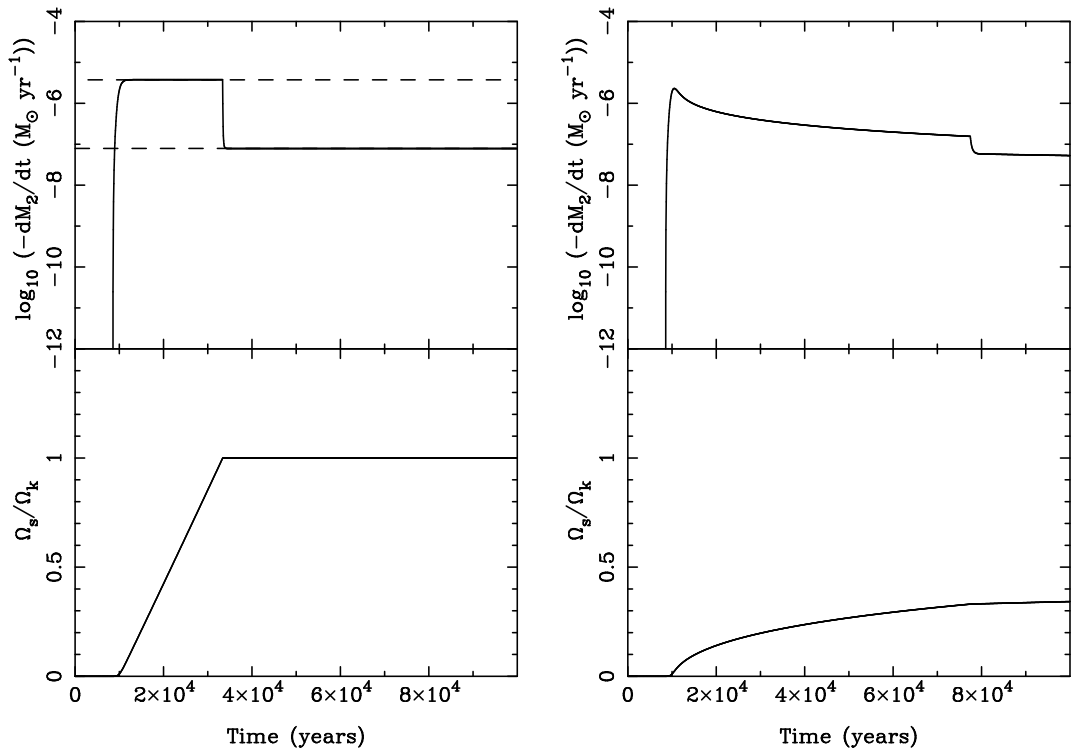}
\hspace*{\fill}
\caption{The mass transfer rate and accretor's spin rate (relative to
the break-up rate) shortly after the start of mass transfer between
white dwarfs of masses $0.5$ and $0.11\,\msun$ with very weak
spin--orbit coupling ($\tau_S = 10^{15}\,\yr$).  In the left-hand
panels the masses were held fixed for comparison with analytic
predictions (dashed lines).  The right-hand panels show that
evolution is strong enough in this case to avoid the break-up limit,
and also to induce a transition to disc-fed accretion after
$80$,$000\,\yr$.}
\label{fig:switchon}
\end{figure*}
For example, Webbink \& Iben (1987) estimate that the field
strength would need to be of order $10^{10}\,\mathrm{G}$ to reduce the
synchronisation timescale to the order of decades. One point worth
making about Campbell's expression is that the
$a^6$ dependence is largely compensated by the $R_2^{-5}$ terms since
$R_2$ will expand with $a$, apart from the relatively small effect of
the changing mass ratio. If this still applies to white dwarfs, then
it means that if magnetic torques stabilise the mass transfer, the
systems will remain almost synchronised at long orbital periods too. 
These systems will appear as ``polars'' with magnetically controlled
accretion.

Tides provide another synchronisation mechanism. In the standard
formalism for tidal synchronisation, tidal deformation in an
asynchronous system is damped by some form of viscosity or radiative
damping (Alexander 1973; Zahn 1977;
Campbell 1984; Eggleton, Kiseleva \& Hut 1998). Radiative damping is
more effective than the viscosity of degenerate matter. Campbell (1984) derives the expression
\begin{equation}
\tau_S = 1.3 \times 10^7 \left(\frac{M_1}{M_2}\right)^2 \left(\frac{a}{R_1}\right)^6 
\left(\frac{M_1/\msun}{L_1/\lsun}\right)^{5/7} \,\yr, 
\label{eq:campbell}
\end{equation}
(modified to reflect synchronisation of the accretor rather than the
donor).  Others differ in detail, but retain the scaling with mass
ratio and orbital separation. In contrast to the magnetic case, the
size of the donor does not enter this expression and thus the tidal
torque drops off rapidly with increasing separation. Unfortunately, as
we discuss later, the overall magnitude of the timescale seems extremely
uncertain. We therefore take it as a free parameter defined by the
timescale at the moment of first contact but retain the scaling with
mass ratio, radius of the accretor and orbital separation of the
above expression, i.e. we assume that 
\begin{equation}
\tau_S \propto \left(\frac{M_1}{M_2}\right)^2 \left(\frac{a}{R_1}\right)^6 .
\end{equation}
We now look at the results of long-term computations of the evolution for
a wide range of initial masses of each component.

\subsection{No evolution versus evolution of parameters}
Before presenting the full results of the numerical integrations,
we pause to compare a numerical calculation with and without evolution
(Fig.~\ref{fig:switchon}) as a test of both the analytic
predictions (details of which can be found in the appendix) and the code.

This figure shows the switch-on of mass transfer between two white dwarfs, for a
stable case satisfying Eq.~\ref{eq:stable_app}, with negligible spin--orbit
coupling. Initially the mass transfer rate rises sharply, but once the mass
transfer rate is high enough to counter-balance the effect of GR, there follow a
few thousand years of steady mass transfer at a relatively high rate, during
which the accretor spins up. The predicted rate from the analysis of the
appendix is marked as a dashed line in the upper-left panel of
Fig.~\ref{fig:switchon}, and matches the mass transfer used by
Nelemans et~al. (2001) for such cases. Once the accretor reaches the break-up
rate, then our assumed stronger coupling sets in, injecting angular momentum
back into the orbit and causing a steep drop in mass transfer rate.

In this particular case ($M_1 = 0.5\,\msun$, $M_2 = 0.11\,\msun$), if
the parameters are allowed to evolve (as described in the next section
and plotted in the right-hand panels of Fig.~\ref{fig:switchon}), then
the break-up spin is not reached within the period of time shown, but
the system makes a transition from direct impact to disc-fed accretion
which causes a similar, although less dramatic, drop in accretion rate
after $80$,$000\,\yr$.

\subsection{Long-term integrations}
\begin{figure*}
\hspace*{\fill}
\includegraphics[angle=270,width=\columnwidth]{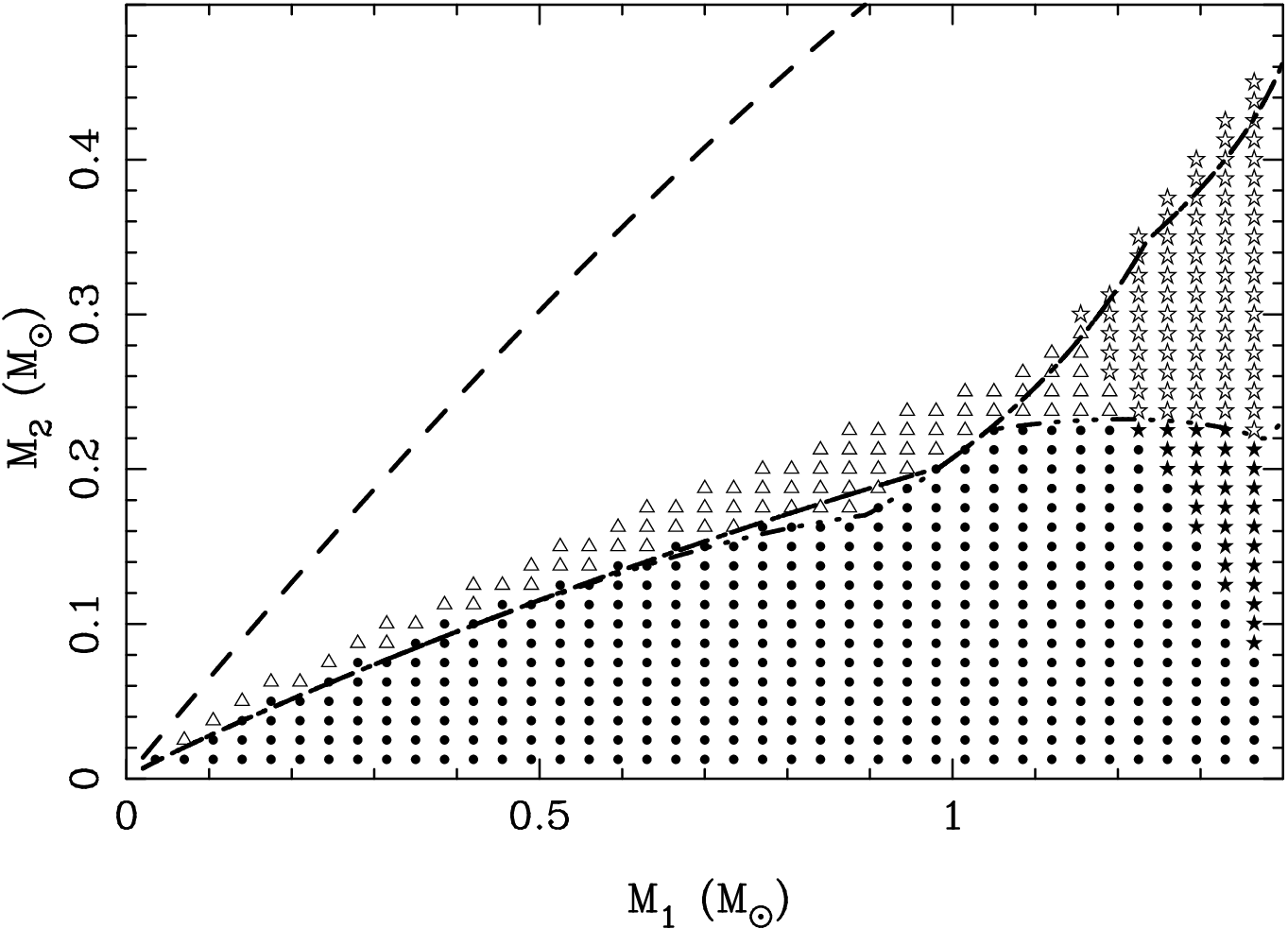}
\hspace*{\fill}
\includegraphics[angle=270,width=\columnwidth]{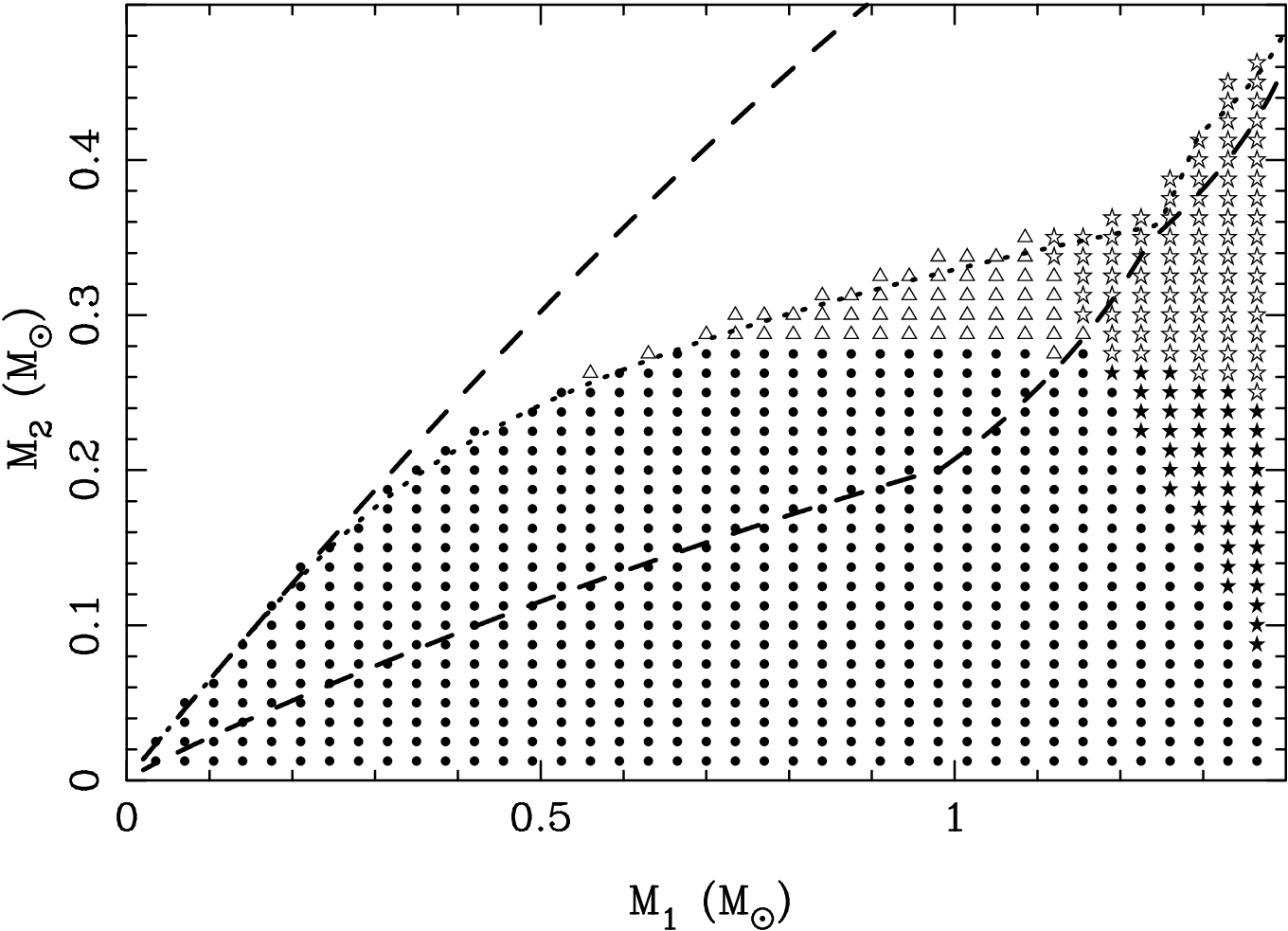}
\hspace*{\fill}
\caption{The left-panel shows the results of evolution for $10^9\,\yr$ following contact
for weak coupling with $\tau_S = 10^{15}\,\yr$. The right-hand panel shows
the case of $\tau_S = 10\,\yr$. In each case, models were computed on a regular grid
covering the whole parameter range shown, but only stable models are
plotted, i.e. all the models in the empty upper-left regions are
dynamically unstable. Open triangles indicate super-Eddington
accretion rate systems; open stars indicate super-Eddington accretion
rate systems with a total mass in excess of the Chandrasekhar limit;
filled stars mark systems in which the accretor reaches the
Chandrasekhar limit within $10^9\,\yr$. We identify the filled circles
(stable, sub-Eddington, sub-Chandrasekhar) as AM~CVn
progenitors. Analytic limits are plotted as in
Fig.~\protect\ref{fig:param} with the addition in the left-hand panel of a
dot--dot--dot--dashed line dividing sub- from super-Eddington systems,
calculated in the weak coupling limit (Eq.~\protect\ref{eq:mdhigh}). 
Note that this line largely overlaps the dashed line marking instability 
except for around $M_1 \approx 0.85\,\msun$ and $M_1 > 1\,\msun$. }
\label{fig:grid_weak_medium}
\end{figure*}

\begin{figure}
\hspace*{\fill}
\includegraphics[angle=270,width=\columnwidth]{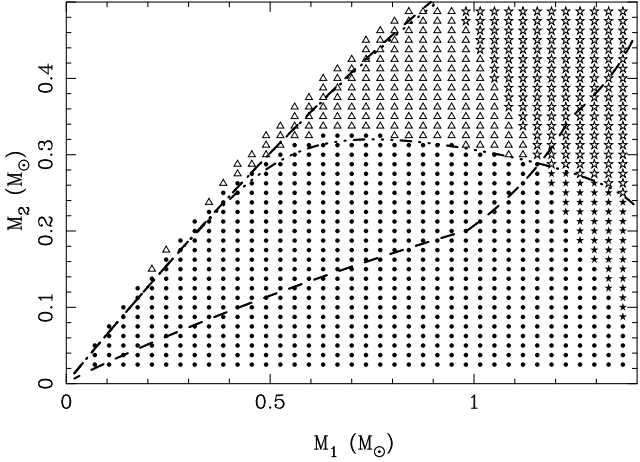}
\hspace*{\fill}
\caption{As for Fig~\protect\ref{fig:grid_weak_medium}, but for very strong
coupling ($\tau_S = 0.1\,\yr$). The dot--dot--dot--dashed line which divides
sub- from super-Eddington systems were computed in the strong coupling limit
(Eq.~\protect\ref{eq:mdnorm}). The dotted line marking the instability limit for
very strong coupling almost coincides with the upper dashed line marking
absolute instability.}
\label{fig:grid_strong}
\end{figure}

With the above scaling of the synchronisation torque, we computed evolution over
a grid in $M_1$, $M_2$ parameter space to determine the long-term stability of
systems. Each model was followed for $10^9\,\yr$ after contact, or until it
became unstable (defined as $|\mtwodot| > 0.01\,\msun\,\yr^{-1}$), or until the
accretor exceeded the Chandrasekhar limit (defined here as $1.438\,\msun$ to
avoid problems when computing the radius of the accretor using Eggleton's
formula). We computed the grid for a number of different synchronisation
timescales at contact. The results for $\tau_S = 10^{15}\,\yr$ and $10\,\yr$ are
shown in Fig.~\ref{fig:grid_weak_medium} while 
Fig.~\ref{fig:grid_strong} shows the case of very strong coupling with
$\tau_S = 0.1\,\yr$. In all cases the points plotted represent the masses of the
stars immediately before the start of mass transfer. In these figures we plot
the analytic stability limits as shown in Fig.~\ref{fig:param}.  Furthermore,
for the $\tau_S = 0.1$ and $\tau_S = 10^{15}\,\yr$ cases we use
Eqs.~\ref{eq:mdnorm} and \ref{eq:mdhigh} respectively to compute the donor mass
above which the accretion rate will be super-Eddington (an iterative
calculation, there is no simple functional form for this limit); these limits
are plotted as dot--dot--dot--dashed lines.

\begin{figure}
\hspace*{\fill}
\includegraphics[angle=270,width=\columnwidth]{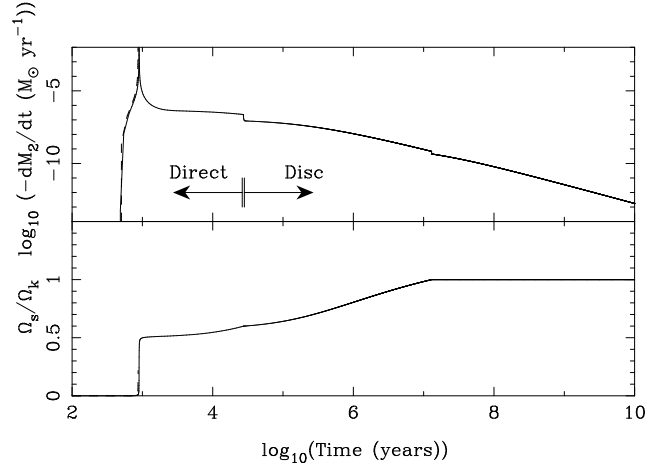}
\hspace*{\fill}
\caption{The long-term evolution of two weakly coupled ($\tau_S =
10^{15}\,\yr$) systems are shown, with one just below the instability limit
($M_1 = 0.5\,\msun$, $M_2 = 0.150\,\msun$, solid line)  and the other just above 
($M_1 = 0.5\,\msun$, $M_2 = 0.151\,\msun$, dashed).}
\label{fig:compare}
\end{figure}

The main result of our integrations is that neither spinning up to break-up nor
the weakening of the synchronisation torque as the system evolves have much
effect, i.e.\ the analytic stability limit Eq.~\ref{eq:sync} provides a fair
estimate of whether a system will survive mass transfer or not. This is because
instability either sets in on a very short timescale, or not at all. Even those
systems which are numerically stable while analytically unstable (i.e. they
violate the analytic stability limit) have little overall effect upon survival
rates because they nearly always exceed the Eddington limit
(Fig.~\ref{fig:grid_weak_medium}).  The small effect that the weakening of the
synchronising torques has can be put down to the concurrent lengthening of the
GR timescale. We illustrate this in Fig.~\ref{fig:compare} which shows the
evolution over $10^{10}\,\yr$ of two weakly coupled systems with the same
mass for the accretor $M_1 = 0.5\,\msun$, but with donor masses of $0.125$ and $0.126\,\msun$,
the first of which is stable in the long term while the second is not.  This
figure shows why our analytic criteria are useful: the important action happens
so fast that complicating factors such as breakup have little effect.

Three major effects distinguish our numerical integrations from the analytical
work. First, the Eddington luminosity is exceeded by some of the dynamically
stable systems. We assume that these will merge.  Super-Eddington systems are
marked in Figs~\ref{fig:grid_weak_medium} and \ref{fig:grid_strong} by
open triangles.  The second effect,
which promotes stability, is the evolution of the stellar masses towards the
region of stability. This can in some cases save a system by moving it from an
analytically unstable position to a stable one before the instability has had
time to set in. However, as remarked above, many of these ``saved'' systems
suffer super-Eddington accretion, so there is little overall increase in
survival rates. Third, and most obvious, is that some systems are able to exceed
the Chandrasekhar limit. While it is not clear \textit{a priori} how long they
will take to do this, it turns out that the mass transfer rate is initially so
high, that the total mass of the binary need only be a very small amount over
the Chandrasekhar limit for this to happen within $10^9\,\yr$, relatively short
compared to the lifetimes of AM~CVn systems.  Therefore these systems, although
of considerable interest as potential Type~Ia supernova progenitors, will be too
short-lived and intrinsically rare (because of the large mass of the primary
star required) to contribute much to the total population of AM~CVn systems (see
Fig.~\ref{fig:birth_params} for example).

Super-Eddington accretion becomes most significant for strong coupling, as seen
in the left-hand panel of Fig.~\ref{fig:grid_strong}. In this panel, the
analytic limit which divides sub- from super-Eddington accretion
(dot--dot--dot--dashed line, calculated from Eq.~\ref{eq:mdnorm}) agrees well
with the numerical integrations.  However, the analytic limit is obtained in the
appendix in the limit of rigid coupling between the accretor and the orbit, so
no more parameter space can be gained for the production of AM~CVn systems by
reducing $\tau_S$ any further. The analytic limit for the case of weak coupling
(Eq.~\ref{eq:mdhigh}) matches the $\tau_S = 10^{15}\,\yr$ case well (left panel
of Fig.~\ref{fig:grid_weak_medium}). However, we were not able to obtain
accurate analytic predictions for the intermediate $\tau_S = 10\,\yr$ case
(right-hand panel of Fig.~\ref{fig:grid_weak_medium}) because in this case it is
not correct to assume either that the accretor is locked as for
Eq.~\ref{eq:mdnorm} or that it is freely rotating as for Eq.~\ref{eq:mdhigh},
and instead one must compute the spin evolution fully.

We have ignored several effects that may be of significance. First, our models
make no allowance for tidal heating during the approach to contact
(Webbink \& Iben 1987; Rieutord \& Bonazzola 1987; Iben, Tutukov \& Fedorova 1998). This might
increase the range of validity of the isothermal mass transfer mode, although it
is hard to see it having a significant effect given that the transition transfer
rate for the isothermal versus adiabatic modes is so low. If
Campbell's (1984) dependence of tidal
synchronisation upon luminosity applies, there could also be a feedback of tidal
heating into a stronger synchronisation torque. This could be modelled, but with
the details of synchronisation so unclear, we prefer just to note that such
effects might occur. While on this issue, it is worth emphasizing that we use a
zero-temperature mass-radius relation for both white dwarfs, whereas we can
expect larger radii for a given mass from finite entropy effects, depending upon
the time taken to reach contact (Deloye \& Bildsten 2003).  It is also worth
pointing out that if tidal heating is not significant, then the donor may not
even be synchronised, which would necessitate changing the whole prescription
based upon Roche geometry. Another issue that we have not tackled is the effect
of helium ignition on the accreting white dwarf (since the donors of interest
for survival of mergers are all low mass and therefore of helium rather than
carbon-oxygen composition). It is not clear whether helium ignition, if it takes
place, would increase or decrease the probability of avoiding the merger.

\begin{figure}
\hspace*{\fill}
\includegraphics[angle=270,width=\columnwidth]{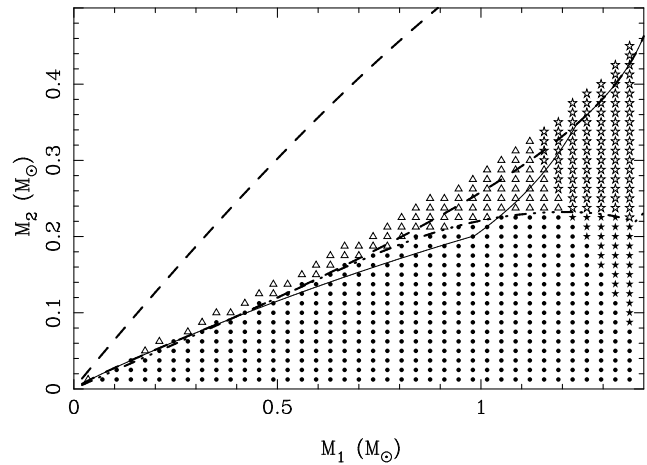}
\hspace*{\fill}
\caption{This figure again shows very weak coupling ($\tau_S = 10^{15}\,\yr$, cf
left-hand panel of Fig~\protect\ref{fig:grid_weak_medium}), but with the
accretion forced to be through a disc. The lower dashed line shows the
guaranteed stability limit for disc accretion, while the thin solid line
shows the usual limit when direct impact is included. }
\label{fig:grid_disc}
\end{figure}

\subsection{Destabilisation of Disc Accretion}
\label{sec:cvs}
It is the finite size of the accretor which leads to the destabilising removal
of angular momentum from the orbit as mass is transferred. Direct impact
accretion is a secondary, although important, consequence of this. To illustrate
this point, we repeated the weak coupling calculations shown in the left-hand
panel of Fig.~\ref{fig:grid_weak_medium} while assuming that all accretion
occurs through a disc, regardless of the whether the minimum stream radius was
smaller than the primary star or not (one could envisage for instance
initialising these systems in a state of disc accretion, however unrealistic
this is in practice). The results are shown in Fig.~\ref{fig:grid_disc}. Rather
surprisingly perhaps, we see that the standard ``$q = 2/3$'' limit, marked by
the upper dashed line, is not much better for disc accretion than it is for
direct impact accretion. This is also clear from the relatively small change
between the thin solid line (direct impact included) and the lower dashed line
(disc accretion only). As we said earlier, this is because the primary star's
radius $R_1$ is not that much smaller than the circularisation radius $R_h$ in
these systems.  For disc accretion, the zero spin--orbit coupling stability
limit of Eq.~\ref{eq:stable} becomes
\begin{equation}
q < 1 + \frac{\zeta_2-\zeta_{r_L}}{2} - \sqrt{(1+q) \frac{R_1}{a}},\label{eq:discstable}
\end{equation}
which is plotted as the lower dashed line in Fig.~\ref{fig:grid_disc}. 

\begin{figure}
\hspace*{\fill}
\includegraphics[angle=270,width=\columnwidth]{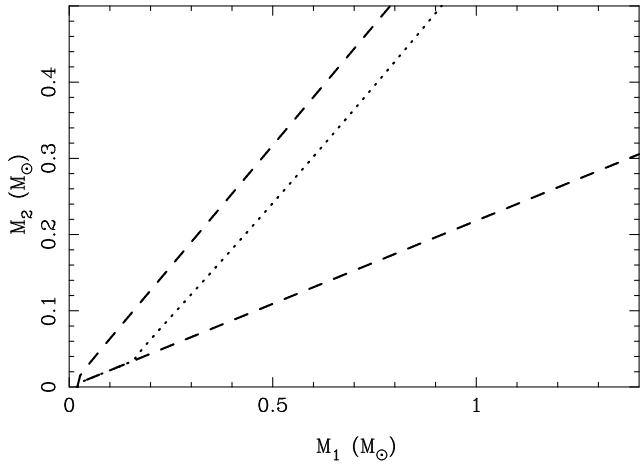}
\hspace*{\fill}
\caption{The regions of stability for cataclysmic variable stars with negligible spin--orbit coupling
($\tau_S = 10^{15}\,\yr$). Systems below the dotted line (Eq.~\protect\ref{eq:discstable}) are stable, 
whereas under the usual assumption that angular momentum is entirely fed back, the upper dashed line 
would be the relevant limit. The lines on this figure are the CV equivalents of those plotted in 
Fig.~\protect\ref{fig:grid_disc}.}
\label{fig:cv_stability}
\end{figure}

This raises the question of whether sinking angular momentum into the accretor
also has a significant effect in cataclysmic variable stars (CVs). To evaluate
this, we computed the analytic stability limit with a revised mass--radius
relation for the donor. We assumed that $R_2 = M_2$ (solar units) and that
$\zeta_2 = -1/3$ (measuring the adiabatic response), approximately correct for
low mass donors. The result is shown in Fig.~\ref{fig:cv_stability}.  As
expected, given the larger orbits of CVs, the destabilisation is nothing like as
severe as it is for double white dwarfs, but neither is it entirely
negligible. The impact of this upon the formation of CVs depends upon the
distribution of pre-CVs in the $(M_1$,$M_2)$ plane, but has some potential to
favour magnetic over non-magnetic systems and to reduce overall formation rates.

\section{Survival as a function of synchronisation}

\label{sec:survive}
\begin{figure}
\hspace*{\fill}
\includegraphics[angle=270,width=\columnwidth]{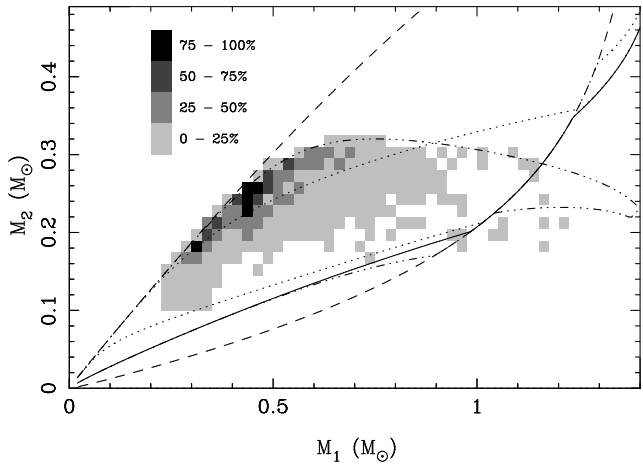}
\hspace*{\fill}
\caption{The masses of double white dwarfs at birth from the binary population
models of Nelemans et~al. (2001) that survive in the case of strong coupling
($\tau_S = 0$).  The surviving models do not reach the limit of guaranteed
stability in this case (upper dashed curve) because super-Eddington accretion
becomes the more stringent constraint (as marked by the dot--dashed
line). Dotted lines are as plotted in Fig.~\protect\ref{fig:param}}
\label{fig:birth_params}
\end{figure}

\begin{figure}
\hspace*{\fill}
\includegraphics[angle=270,width=\columnwidth]{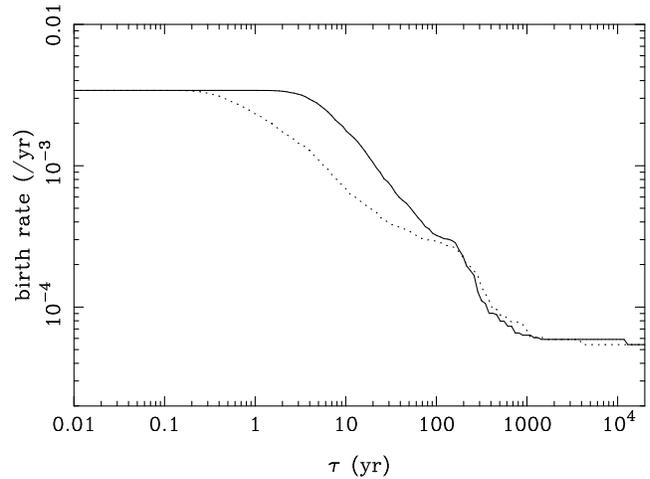}
\hspace*{\fill}
\caption{The birth rate of AM~CVn systems descended from double white dwarfs as
a function of the synchronisation timescale at the start of mass transfer. We
show also the case for isothermal mass transfer (dotted line) to show that the mass
transfer mode has a relatively slight overall effect. The plot was derived from
our analytic approximations for speed, but limited comparisons with numerical
integrations produced nearly-identical results.}
\label{fig:birth_rate}
\end{figure}

We are unlikely to catch a system in the short-lived phase after the
start of mass-transfer, and so the most important consequence of the
destabilising effect of finite accretor size is its impact upon the
survival of double white dwarfs as AM~CVn systems. To gain an idea of
how significant this could be, we applied our stability criterion to
the models of Nelemans et~al. (2001) to obtain the predicted birth rate
of AM~CVn stars as a function of the synchronisation timescale at the
start of mass transfer. Fig.~\ref{fig:birth_params} shows a greyscale
representation of the stellar masses at birth from the models.
The greyscale only shows systems that can become AM~CVn
binaries. They concentrate towards the line dividing sub- from
super-Eddington systems because they are outliers from the majority of
systems which have near-equal mass ratios. This depends upon the
physics of the common envelope amongst other things:
Nelemans et~al. (2001) developed their model partly in response to the
observations of equal mass ratios of double white dwarfs. Others have
found more unequal distributions (Iben et~al. 1997; Han 1998),
nevertheless all models so far peak with $q > 0.4$ and thus all
predict that most systems are unstable and will merge. The high density near the
stability limit of Nelemans et~al.'s distribution means
that the birth rate of AM~CVn systems from the double white dwarf path
is very sensitive to the strength of synchronisation torque.  This can
be seen in Fig.~\ref{fig:birth_rate}
in which we plot the birth-rate (from the double white dwarf merger route) 
as a function of the synchronisation timescale \footnote{The extreme values in
Fig.~\protect\ref{fig:birth_rate} differ from
those of models I and II in Nelemans et~al. (2001) because we use a different mass--radius
relation in the calculation of the Eddington rate.}.  This figure shows that as 
spin--orbit coupling weakens, the birth rate of AM~CVn stars from double white dwarfs (which we
assume to be equivalent to survival of mass transfer as a binary)
drops by about one hundred-fold. A synchronisation timescale $\tau_S <
1000\,\yr$ is enough to increase the birth rate from its lowest value
by more than a factor of two. While sensitive to the mass ratio
distribution of the detached double white dwarfs, this calculation
illustrates the potential significance of dissipative synchronisation
torques for the population of AM~CVn systems. In this figure we also show the results
for isothermal mass transfer, which shows that variations between the two extremes
is not that large.

One could hope that an observational estimate of the birth rate of AM~CVn stars
might decide whether we are in the strong or weak coupling part of
Fig.~\ref{fig:birth_rate}. Unfortunately in our opinion, uncertainties and
selection effects are too large to allow this to be done.  First, there are only
two directly-measured distances (GP~Com, $68 \pm 7 \,{\rm pc}$,
Thorstensen 2004; AM~CVn, $235\,{\rm pc}$, C.Dahn\footnote{On behalf
  of the USNO CCD parallax team}, priv.\ comm.). The AM~CVn distance is a factor
of 3 further than Warner (1995) assumed in estimating the space density, which
if it applied to all systems would imply an overestimate by a factor of 27,
which serves to show how uncertain observational estimates are.  To add to this,
Nelemans et~al. (2001) estimate that while 99\% of AM~CVns have periods $P >
2000\,{\rm sec}$, they will only form $\sim 10$\% of the observed sample, an
estimate that itself depends upon highly uncertain assumptions about selection.
Finally, the models that lead to Fig.~\ref{fig:birth_rate} have uncertain
normalisation -- a factor of 5 to 10 seems quite possible. There is clearly
considerable scope for work in this area.

\section{Discussion}
\label{sec:discuss}
The requirement that the initial synchronisation timescale is below $1000\,\yr$
in order to raise survival rates of double white dwarfs as binary systems is the
most important result of this paper.  There are few calculations of
synchronisation of white dwarfs in binaries, but those that exist come nowhere
near this strength of synchronisation. For instance, Campbell's
(1984) estimate for radiative damping (Eq.~\ref{eq:campbell})
applied to a pair of white dwarfs with masses of $M_1 = 0.5\,\msun$ and $M_2 =
0.2\,\msun$ gives $\tau_S \sim 10^{12}\,\yr$.  Even longer times
($10^{15}\,\yr$, Webbink \& Iben 1987) emerge from calculations based upon the
viscosity of degenerate matter.  Indeed, the estimates of synchronisation
timescales are so long that even the mass donor would not have achieved
synchronism prior to mass transfer let alone the accretor
(Mochkovitch \& Livio 1989).  If these estimates are reliable, then the lowest
birth rate estimates of Nelemans et~al. (2001) are to be preferred.  However, it
is widely recognised by the authors of the papers on synchronisation of white
dwarfs that the problem is essentially unsolved: the possibilities of turbulent
viscosity (Horedt 1975) and the excitation of non-radial modes
(Campbell 1984) can dramatically shorten the synchronisation time. For
instance, Mochkovitch \& Livio (1989) argue that turbulent viscosity can give
synchronisation timescales $\ll 500\,\yr$, exactly what is needed to raise the
birth rates significantly. The uncertainty over the dissipation mechanisms
within white dwarfs is thus a major unsolved problem for AM~CVn evolution.

If the synchronisation torques are not strong, then either the ultra-short
period systems (if they are such) are descended from systems that have always
had extreme enough mass ratios to be stable, or the double white dwarf route may
not be the main channel for these systems. The former alternative is possible,
despite the absence of a single system of extreme mass ratio amongst the
observed close double white dwarf population (Maxted et~al. 2002),
because they are expected to be rare, as shown by Fig.~\ref{fig:birth_params}.
This would mean that AM~CVn birth rates through the double white dwarf route are
dependent upon on a so-far-untested part of the mass ratio distribution of
double white dwarfs. The main objection to formation other than from two white
dwarfs is the relative difficulty of reaching ultra-short orbital periods by
other routes: it is important from this point of view that the photometric
periods of order 10 minutes or less are confirmed (or not) as orbital periods.

One way to test our ideas would be to establish that there are systems in
existence that are stabilised by synchronisation torques. Although such systems
are likely to be in a phase of direct impact accretion, it is important to
realise that direct impact on its own does \emph{not} equal instability.  There
is a region of parameter space of stable, direct impact accretion even in the
absence of synchronisation torques (the slim lenticular region in
Fig.~\ref{fig:param} containing the word ``Direct''). For instance, V407~Vul, if
it is a direct impact accretor, could be in this region (Marsh \& Steeghs 2002);
so too could RXJ~J0806.3+1527. Thus even if direct impact accretion were to be
established in a double white binary, it would not mean that significant
synchronising torques were acting.  We would need in addition to establish that
the system parameters implied instability in the absence of such torques.  Short
period systems are of particular interest in this respect since short periods
require high donor masses, and correspondingly high accretor masses if they are
to remain in the lenticular region. This is an excellent reason for attempting
to find further examples of such systems. X-ray surveys would seem the most
promising method, since both V407~Vul and RX~J0806.3+1527 were found in this
manner, although in the future perhaps space-based gravitational wave detectors
will prove even more useful (Nelemans, Yungelson \& Portegies Zwart 2004).

Similar remarks apply to CVs, except for these stars the unstable region of
Fig.~\ref{fig:cv_stability} is small enough that it seems unlikely that the
stellar masses can be measured with enough accuracy to be certain that a given
system is located within it.

\subsection{Measuring synchronisation torques from other binary stars}

The strength of synchronisation torques on white dwarfs might be testable in
other systems. If any short period systems with rapidly rotating white dwarfs
can be identified, it might be possible to place a lower limit upon the
synchronisation torque acting. For instance, rotation velocities have been
measured for several white dwarfs in CVs (Sion 1999). The strongest
constraints come from high spin-rate, low accretion-rate, short-period
systems. Perhaps the best example is the cataclysmic variable WZ~Sge, which has
an orbital period of $81.6\,\min$ and an accretor which shows signs of having a
spin period of $\approx 28\,\sec$ (Patterson et~al. 1998). WZ~Sge is an old, low
accretion rate system, accreting via a disc. Assuming that its spin period
represents the equilibrium between accretion and tidal dissipation torques, and
since $\Omega_s \gg \Omega_o$, we have
\begin{equation}
\tau_S = \frac{k M_1 R_1^2 \Omega_s}{|\mtwodot| \sqrt{GM_1 R_a}},
\end{equation}
where $R_a$ is the radius equivalent to the specific angular momentum of the
accreted material. We expect $R_a$ to lie in the range $R_1 < R_a < R_h$
(allowing $R_a \neq R_1$ because the accretor may be magnetised given the
presence of a spin signal).  We take for the mass transfer rate $|\mtwodot| = 3
\times 10^{-11}\,\msun\,\yr^{-1}$, estimated from Fig.~1 of Kolb \& Baraffe (1999)
as we are interested in the long term average value since the timescale turns
out to be long.  Thus assuming for WZ~Sge that $M_1 = 1\,\msun$ and $M_2 =
0.08\,\msun$ (Steeghs et~al. 2001), we find
\begin{equation}
3 \times 10^8 < \tau_S < 1.4 \times 10^9 \,\yr .
\end{equation}
Armed with the constraint on $\tau_S$ from WZ~Sge, we can scale to a double
white dwarf with $M_1 = 1\,\msun$ and $M_2 = 0.26\,\msun$.  The separation of
such a system is $9.5$ times smaller than WZ~Sge, while the donor is $3.25$
times more massive.  Scaling according to $\tau_S \propto a^6 M_2^{-2}$ ($M_1$
and $R_1$ being the same), the equivalent range of $\tau_S$ for the double white
dwarf is
\begin{equation}
40 < \tau_S < 180 \,\yr ,
\end{equation}
very much the sort of values that can have an impact upon AM~CVn evolution.

Unfortunately, there are significant problems with the above analysis.  If the
pulsation in WZ~Sge really does mean that the accretor is magnetic (and truly
represents the white dwarf's spin period), then it may be magnetic coupling to
the outer accretion disc which keeps it rotating slowly. In addition there are
complications of nova outbursts in CVs which probably make them unsuitable for
measurement of $\tau_S$ under any circumstances (Livio \& Pringle 1998). Detached
systems avoid these problems, and perhaps might make it possible to measure a
spin-down rate if a weakly-magnetic example could be found (but strong enough to
give a spin pulse) and it could be shown that magnetic and accretion torques
were negligible. Interestingly there is apparently a detached system which
contains a fairly rapid rotator (EC13471-1258, O'Donoghue et~al. 2003), but
in this case it is suggested that this was caused by a recent episode of
accretion, and so it is not an indication of weak synchronisation. We are left
with no clear answer as to the strength of synchronising torques from the known
types of white dwarf binary stars. Nevertheless, it is clear from the estimate
above that if tidal synchronisation is significant for the evolution of AM~CVn
systems, then it should also be important in fixing the spin rates of white
dwarfs in cataclysmic variable stars and similar systems, and this could be a
possible explanation for the generally low rotational velocities detected so far
in cataclysmic variable stars, none of which are near the break-up rate
(Sion 1999).

\subsection{Magnetic versus tidal synchronisation}

If magnetic rather than tidal torques were dominant, an interesting filtering
effect could occur in which systems with strongly magnetic accretors survive
mass transfer while non-magnetic systems mostly merge. One would then expect
magnetic accretors to be over-represented amongst the AM~CVn systems. This is
not obviously the case, e.g. GP~Com has no detectable circular polarisation
(Cropper 1986) and along with V396~Hya shows the clear signature of
emission from a disc, while other systems in the class show shallow absorption
line spectra similar to those of high-state non-magnetic CVs rather than the
magnetic AM~Her class. This suggests that either the fraction of merging double
white dwarfs that are magnetic is so low that even the pruning of 99 per cent of
the non-magnetic systems fails to reveal them, or that tidal synchronisation is
effective in stabilising both non-magnetic and magnetic systems, or that neither
magnetic nor tidal synchronisation has much effect or, finally, that AM~CVn
stars do not form through the double white dwarf route.

\subsection{Period changes}

As several investigators have realised, measurements of period changes have some
potential to discriminate between models of the ultra-compact binary stars. Once
a system has settled onto its long term evolution towards the AM~CVn phase and
spin equilibrium has been reached, it is straight-forward to show that the
period derivative in a semi-detached state $\dot{P}_{SD}$ relative to its pure
GR-driven value when detached $\dot{P}_D$, is given by
\begin{equation}
\frac{\dot{P}_{SD}}{\dot{P}_D} = \frac{(\zeta_2 - \zeta_{r_L})/2}{1 + (\zeta_2 - \zeta_{r_L})/2 - q}. 
\end{equation}
Since $\zeta_{r_L} \approx 1/3$ and $\zeta_2 \approx -1/3$, and since the
denominator of the above expression must be positive for stability, this ratio
is negative; see Table~\ref{tab:pdot} for some explicit values for systems which
are stable even when there are no synchronisation torques.
\begin{table}
\hspace*{\fill}
\begin{tabular}{cccccc}
\hline
$M_1$  & $M_2$   & $\dot{P}_{SD}/\dot{P}_{D}$ & $M_1$  & $M_2$   & $\dot{P}_{SD}/\dot{P}_{D}$ \\
$\msun$& $\msun$ &                            & $\msun$& $\msun$ & \\
\hline
0.3    & 0.06    & -0.61 & 0.7    & 0.08    & -0.53 \\
0.5    & 0.06    & -0.52 & 0.7    & 0.10    & -0.58 \\
0.5    & 0.08    & -0.58 & 0.7    & 0.12    & -0.64 \\
0.5    & 0.10    & -0.66 & 0.7    & 0.14    & -0.70 \\
0.7    & 0.06    & -0.48 &        &         &\\
\hline
\end{tabular}
\hspace*{\fill}

\caption{The ratio of the period derivative in a semi-detached state
to its value when (just) detached is given for different stellar masses.}

\label{tab:pdot}

\end{table}
To date period changes have been reported for V407~Vul
(Strohmayer 2002) and RX~J0806.3+1527 (Hakala et~al. 2003;
Strohmayer 2003), and in each case the period was found to be
decreasing. These results favour a detached rather than semi-detached
configuration for these two systems (ES~Cet on the other hand is clearly
accreting). This would weaken the case for V407~Vul and RX~J0806.3+1527 having
passed through the state discussed in this paper, although period changes in
other close binary stars such as the cataclysmic variable stars have proved
unreliable as a means of measuring long-term angular momentum changes
(Applegate 1992; Baptista et~al. 2003). The periods of the
ultra-compact systems should continue to be monitored to see whether similar
problems affect these stars, although one might hope that at such short periods
gravitational wave angular momentum losses will dominate over any secondary
effects.

\section{Conclusions}

We have studied the onset of mass transfer between two white dwarfs, in
particular its stability during the phase when the accretion stream hits the
accretor directly. We find that this phase can be stabilised by coupling of the
spin of the accretor to the binary orbit, through dissipative processes such as
tidal stressing and magnetic induction.  However, the coupling needs to be
strong and must act on a timescale $< 1000\,\yr$ when mass transfer first starts
to have much effect upon the survival rates of the systems. Standard estimates
of the synchronisation timescales in white dwarfs are orders of magnitude longer
than this, but may wildly underestimate the synchronisation torques. We have
also found that during disc accretion, the angular momentum lost to the accretor
at the inner edge of disc also destabilises mass transfer, and is almost as
significant as the direct impact case, essentially because the radius of the
accretor is comparable to the circularisation radius in these systems.  The same
effect plays a lesser but non-negligible role in non-magnetic cataclysmic
variable stars. Further discoveries of ultra-compact double white dwarfs have
the potential to tell us whether synchronisation torques are indeed important
for surviving this exciting phase of evolution.

\section*{Acknowledgements}
TRM thanks the Institute of Astronomy, Cambridge, for their hospitality during
the visit when this work was started, and acknowledges financial support of a
PPARC SRF. GN acknowledges the financial support of PPARC and thanks Mike
Montgomery for useful discussions; DS acknowledges the support of a PPARC
fellowship and a Smithsonian Astrophysical Observatory Clay Fellowship. We thank
the referee, Ron Webbink, for pointing out the importance of adiabatic mass
transfer, and for other useful comments. This research has made use of NASA's
Astrophysics Data System Bibliographic Services.

\appendix

\section{Quasi-static solutions}

In this appendix we are interested in quasi-static solutions of
equations~\ref{eq:deldot} and \ref{eq:omdot}, ``static'' on timescales short
compared to the secular evolution of the system parameters and their
stability. As outlined in section~\ref{sec:equilibrium}, we assume therefore
that the coefficients involving binary masses and separation are constant.

We define the following dimensionless variables
\begin{eqnarray} 
t &\rightarrow&
t/\tau_S, \\ x &=& -\tau_S \frac{\mtwodot}{M_2},\\ y &=&
\frac{\omega}{\Omega_o}, \\ z &=& \frac{\Delta}{2R_2}, 
\end{eqnarray} 
the first being a rescaling of the timescale. With these new variables, 
equations~\ref{eq:deldot} and \ref{eq:omdot}, can be written as
\begin{eqnarray} 
\frac{d z}{d t} &=& a_1 - a_2 y - a_3 x \label{eq:dim1}\\ 
\frac{d y}{d t} &=& - a_4 x y - y + a_5 x, \label{eq:dim2} 
 \end{eqnarray} 
where the dimensionless coefficients $a_1$ to $a_5$ are given by 
\begin{eqnarray} 
a_1 &=& \frac{\tau_S}{\tau_G}, \\ 
a_2 &=& \frac{k r_1^2 (1+q)}{q} ,\\ 
a_3 &=& 1 + \frac{\zeta_2-\zeta_{r_L}}{2} - q - \sqrt{(1+q)r_h} ,\\ 
a_4 &=& \lambda q,\\ 
a_5 &=& \frac{q \sqrt{(1+q)r_h}}{k r_1^2 (1+q)} - \lambda q , 
\end{eqnarray} 
and where $\tau_G = - J_\mathrm{orb}/\dot{J}_\mathrm{GR}$ and $r_1 =
R_1/a$ as before. The coefficients $a_1$, $a_2$ and $a_5$ are positive, while 
$a_3$ and $a_4$ can be either positive or negative. For Eggleton's
mass-radius relation, Eq.~\ref{eq:egg_mr}, for instance, $a_4 < 0$ for $M_1 >
0.40 \,\msun$.

For the solutions of interest, $\dot{y} = \dot{z} = 0$, and we are left with
two simultaneous equations 
\begin{eqnarray} 
a_3 x_e + a_2 y_e &=& a_1, \label{eq:first}\\ 
a_5 x_e - y_e &=& a_4 x_e y_e . \label{eq:second}
\end{eqnarray} 
These lead to a quadratic equation with solutions 
\begin{equation} 
x_e = \frac{-b \pm \sqrt{b^2 + 4 a_1 a_3 a_4}}{2 a_3 a_4}, 
\label{eq:qsols}
\end{equation} 
where 
\begin{equation} 
b = a_3 + a_2 a_5 - a_1 a_4 .  
\end{equation} 
Solutions of physical interest must have real, positive $x_e$, and, for reasons
of stability, if there are two such solutions, the smaller of the two is
favoured.

This divides parameter space into two parts according to whether there are or
are not any solutions. The region where there are solutions is further
sub-divided according to whether or not the solutions are stable.

\subsection{The dynamically unstable case}
\label{sec:unstable}

We know that $a_1 > 0$. Hence, if $a_4 > 0$ and $a_3 + a_2 a_5 < 0$ (which
since $a_2$ and $a_5$ are both positive implies $a_3 < 0$), then $b < 0$ and
$|b^2 + 4a_1 a_3 a_4 | < |b|$, and so there are \emph{no} real and positive solutions for $x_e$.  
The condition $a_3 + a_2 a_5 < 0$ is equivalent to:
\begin{equation} 
q > 1 + \frac{\zeta_2-\zeta_{r_L}}{2} - k r_1^2 (1+q) \lambda ,
\label{eq:unstable_app}
\end{equation}
which was quoted in the main text as Eq.~\ref{eq:unstable}.

If $a_4 < 0$, as is possible, there is formally a positive solution of
Eq.~\ref{eq:qsols}, but only at such high spin rates that we run into the
break-up limit of the accretor. The consequence is that
Eq.~\ref{eq:unstable_app} remains a good condition for instability.

\subsection{The dynamically stable case}
\label{sec:stable}

The reverse of the unstable case is when $a_3 > 0$, when one can show that there
is \emph{always} a solution for real, positive $x_e$, for any $\tau_S$ and
$a_4$. The condition $a_3 > 0$ means that
\begin{equation}
q < 1 + \frac{\zeta_2-\zeta_{r_L}}{2} - \sqrt{(1+q) r_h}, \label{eq:stable_app}
\end{equation}
as derived by Nelemans et~al. (2001). As noted in the main text, this can be adapted
for disc-fed accretion by replacing $r_h$ by $r_1 = R_1/a$.

Systems evolve towards this solution with time since their mass ratios
continuously decrease. We will prove below that the equilibrium is stable in
this case. Before doing so, we look at the equilibrium mass transfer rate in
this stable case. The equations derived here are those used in the comparison of
analytical and numerical results shown in Fig.~\ref{fig:switchon}.

There are two extreme cases as $\tau_S \rightarrow 0$ or $\tau_S \rightarrow
\infty$. As $\tau_S \rightarrow 0$, $a_1 \rightarrow 0$ and
\begin{equation}
 x_e \rightarrow \frac{a_1}{a_3 + a_2 a_5}.
\end{equation}
Dividing out the $\tau_S$ in the definitions of $x$ and $a_1$, then
we have
\begin{equation}
\frac{-\mtwodote}{M_2} \longrightarrow
\frac{-\dot{J}_\mathrm{GR}/J_\mathrm{orb}}{1+(\zeta_2-\zeta_{r_L})/2-q-k r_1^2 \lambda (1+q)}
\label{eq:mdnorm}
\end{equation}
This is, as expected, the usual expression for stable mass transfer, albeit
slightly modified by the term including $k r_1^2$ as a result of the angular
momentum removed from the orbit by the accreting white dwarf owing to its
increasing mass. This term contributes at most $\sim 0.01$, and can safely be
ignored.

When $\tau_S \rightarrow \infty$, we run into break-up of the accretor once
more. It is then clear from Eq.~\ref{eq:deldotalt} that the equilibrium mass
transfer rate is given by
\begin{equation}
\frac{-\mtwodote}{M_2} = 
 \frac{-\dot{J}_\mathrm{GR}/J_\mathrm{orb}}
{1 + (\zeta_2-\zeta_{r_L})/2 - q - k \lambda \sqrt{(1+q) r_1} }.
\label{eq:mdlow}
\end{equation}
The final factor in the denominator is again fairly small, and so this mass
transfer rate is very similar to that of Eq.~\ref{eq:mdnorm}. Thus we conclude
that systems undergoing direct impact accretion which satisfy
Nelemans et~al.'s (2001) strict stability
condition, Eq.~\ref{eq:stable}, always have equilibria with mass transfer rates
close to that of Eq.~\ref{eq:mdnorm}.

Nelemans et~al. (2001) use a different equation to compute the Eddington-limited
mass transfer rate in the stable direct impact case. On their Fig.~1 they plot a
dashed line obtained from the relation
\begin{equation}
\frac{-\mtwodote}{M_2} =
\frac{-\dot{J}_\mathrm{GR}/J_\mathrm{orb}}{1 + (\zeta_2-\zeta_{r_L})/2 - q - \sqrt{(1+q) r_h }}
\label{eq:mdhigh}
\end{equation}
which gives substantially higher mass transfer rates than either of
Eqs~\ref{eq:mdnorm} and \ref{eq:mdlow}.  Although this is not the equilibrium
rate, Nelemans et~al. were in fact correct to use this, at least in
the weak coupling case, as can be seen by considering the situation that applies
as $\tau_S \rightarrow \infty$ while $\omega$ remains finite.
Eq.~\ref{eq:deldot} then becomes
\begin{equation}
\frac{1}{2R_2}\frac{d \Delta}{d t} = 
-\frac{\dot{J}_\mathrm{GR}}{J_\mathrm{orb}}
+\left(1 + \frac{\zeta_2-\zeta_{r_L}}{2} - q - \sqrt{(1+q)r_h} \right) 
\frac{\mtwodot}{M_2},
\end{equation} 
and the equilibrium solution, $\dot{\Delta} = 0$, gives
Eq.~\ref{eq:mdhigh}. Although this a temporary state that will stop once the
white dwarf has spun up to break-up, it is still long-lasting (see
Fig.~\ref{fig:switchon}).  The spin-up phase will last a time $\tau_B$ given by
\begin{equation}
\frac{\tau_B}{\tau_G} \approx k \sqrt{r_1/r_h} 
(1 + (\zeta_2-\zeta_{r_L})/2 - q - \sqrt{(1+q) r_h }),
\end{equation}
which can be long enough for the system to switch to disc accretion, although
once break-up is reached, the mass transfer rate is still given by
Eq.~\ref{eq:mdlow}. These results are the ones illustrated graphically in
Fig.~\ref{fig:switchon}. As spin--orbit coupling increases we expect a peak rate
intermediate between that given by Eq.~\ref{eq:mdlow} and that given by
Eq.~\ref{eq:mdhigh}.

\subsection{Stability of Equilibrium}
The case of most interest to us is when the mass ratio lies in between the
unstable and stable limits of Eqs~\ref{eq:unstable_app} and \ref{eq:stable_app}
when $- a_2 a_5 < a_3 < 0$. In this case synchronisation can bring about an
equilibrium solution which disappears as $\tau_S \rightarrow \infty$.  This then
sets an upper limit on $\tau_S$ required for the existence of an equilibrium
solution. However it turns out that an even stricter constraint comes from the
requirement that the equilibrium solution be \emph{stable}.  This is a new
feature: so far there either have been or there have not been equilibrium
solutions, and if they existed we have implicitly assumed that they were stable
equilibria. We now come across equilibrium solutions that can be stable or
unstable.

Consider the following small perturbations from equilibrium:
\begin{eqnarray}
y &=& y_e  + y', \\
z &=& z_e  + z' , \\
x &=& x_e + \beta x_e z' , 
\end{eqnarray}
where in the last equation we recognize that $x$ varies with $z$ alone since the mass
transfer rate depends explicitly upon $\Delta$ but not $\omega$, and the dimensionless factor
$\beta$ is defined by
\begin{equation}
\beta = 2 R_2\frac{d \log (-\mtwodot)}{d \Delta}.
\end{equation}
Since the mass transfer rate increases monotonically with $\Delta$, $\beta$ is positive.
For the two mass loss models we consider in section~\ref{sec:mdot}
\begin{equation}
\beta_i = 2 \frac{R_2}{H} ,
\end{equation}
for the isothermal case of Eq.~\ref{eq:mdot}, and
\begin{equation}
\beta_a = 6 \frac{R_2}{\Delta_e},
\end{equation}
for the adiabatic case of Eq.~\ref{eq:mdot_ad}, where $\Delta_e$ is the equilibrium value of
the overfill factor; it is the latter expression that we employed throughout the paper.

Applying the perturbation to Eqs~\ref{eq:dim1} and \ref{eq:dim2} while
keeping terms to first order in the small dashed quantities gives
\begin{eqnarray}
\dot{z'} &=& - \beta x_e a_3 z' - a_2 y', \label{eq:perturb1}\\
\dot{y'} &=& \beta x_e (a_5-a_4 y_e) z'  - (1+a_4 x_e) y'.\label{eq:perturb2}
\end{eqnarray}
These are two first-order, coupled equations. Assuming solutions of the form
$e^{p t}$, one obtains an eigenvalue equation for $p$ with
values that are solutions of the following quadratic:
\begin{equation}   
p^2 + \left(1 + a_4 x_e + \beta x_e a_3 \right) p + 
\beta \left( a_1 + a_3 a_4 x_e^2 \right) = 0.
\end{equation}
after using Eqs~\ref{eq:first} and \ref{eq:second} to simplify the third term.
The solutions are stable provided that $\mathrm{Re}(p) < 0$ for both roots. This is
true if the two coefficients of the quadratic are both positive.
The coefficient of the term linear in $p$ is positive if 
\begin{equation}
\frac{1}{\tau_S} > - \left[a_3 \beta + a_4 \right] \frac{-\mtwodote}{M_2} \label{eq:discuss},
\end{equation}
or, when written out in full,
\begin{eqnarray}  
\frac{1}{\tau_S} &>& - \left[
\left(1 + \frac{\zeta_2-\zeta_{r_L}}{2} - q - \sqrt{(1+q)r_h} \right) \beta \right.
 \label{eq:sync_app} \\
& & \;\;\;\;\;\; + \left. q \lambda \rule{0mm}{4mm}\right] \frac{-\mtwodote}{M_2} ,
\nonumber
\end{eqnarray}
which was quoted in the main text (Eq.~\ref{eq:sync}).
It turns out that the coefficient of the constant term is also
positive when this condition is satisfied, so no further restriction
is necessary. Eq.~\ref{eq:sync_app} is a generalisation
of Nelemans et al.'s (2001) strict condition for stability
(Eq.~\ref{eq:stable}) and is the key result of this paper. 
Eq.~\ref{eq:sync_app}, derived from the condition, $a_3 \beta + a_4 > 0$, does
not quite match the criterion for stability regardless of $\tau_S$, Eq.~\ref{eq:stable_app},
which corresponds to $a_3 > 0$. This is because Eq.~\ref{eq:sync_app} permits arbitrarily
high rates of spin, whereas in reality the spin rate is limited by break-up. Both criteria
are correct, but under different conditions: Eq.~\ref{eq:stable_app} applies in the case of 
zero synchronisation, while Eq.~\ref{eq:sync_app} applies when there are synchronising torques 
strong enough to keep the spin rate of the accretor below the break-up rate.

\label{lastpage}

\end{document}